\documentclass[manuscript]{aastex}

\shorttitle{Rapid Oscillations in Cataclysmic Variables}
\shortauthors{Brian Warner}

\begin{document}

\title{Rapid Oscillations in Cataclysmic Variables}

\author{Brian Warner}
\affil{Department of Astronomy, University of Cape Town, Rondebosch, 7700
South Africa}
\email{warner@physci.uct.ac.za}

\begin{abstract}
    I give an overview of the rich phenomenology of dwarf nova 
oscillations (DNOs) and Quasi-periodic Oscillations (QPOs) 
observed in cataclysmic variable stars (CVs). The favoured 
interpretation of these rapid brightness modulations (3 -- $>$1000 s 
time scales) is that they are magnetic in nature -- magnetically 
channelled accretion from the inner accretion disc for DNOs and 
possible magnetically excited travelling waves in the disc for 
QPOs. There is increasing evidence for the magnetic aspects, 
which extend to lower fields the well known properties of strong 
field (polars) and intermediate strength field (intermediate polars) 
CVs . The result is that almost all CVs show the presence of 
magnetic fields on their white dwarf primaries -- though for many 
the intrinsic field may be locally enhanced by the accretion process 
itself.
   
There are many parallel behaviours with the QPOs seen in X-Ray binaries, 
with high and low frequency X-Ray QPOs 
resembling respectively the DNOs and QPOs in CVs.
\end{abstract}

\keywords{cataclysmic variables, stars: binaries: close}

\section{INTRODUCTION}

    The discovery by Merle Walker (1956) on 9 July 1954 of a 71 s 
modulation in the light curve of DQ Her, the remnant of nova 
Herculis 1934, gave notice that cataclysmic variables (CVs) not 
only are binaries of short orbital period, they also are capable of 
producing periodic phenomena on what, in 1954, was of 
unprecedentedly short time scales. The rapidity and high stability 
of the modulation in DQ Her is now attributed to rotation of the 
white dwarf primary, and was the forerunner of the class of CVs 
known as intermediate polars (IPs), in which the magnetic field of 
the primary channels gas from the inner edge of an accretion disc 
onto spots or arcs on the primary, and of which about 32 are now 
known (Kuulkers et al.~2003). In the case of DQ Her the optical 
oscillations have been shown to be due to anisotropic radiation 
from the primary (located probably at two accretion hotspots, so 
the rotation period is actually 142 s), sweeping across the accretion 
disc, deduced from observed phase changes in eclipse of the 
continuum modulation and from the wavelength-dependence of 
pulsation phase in the emission lines (Warner et al.~1972; Patterson, 
Robinson \& Nather 1978; Chanan, Nelson \& Margon 1978; 
Martell et al.~1995; Silber et al.~1996).

\begin{figure}
\plotone{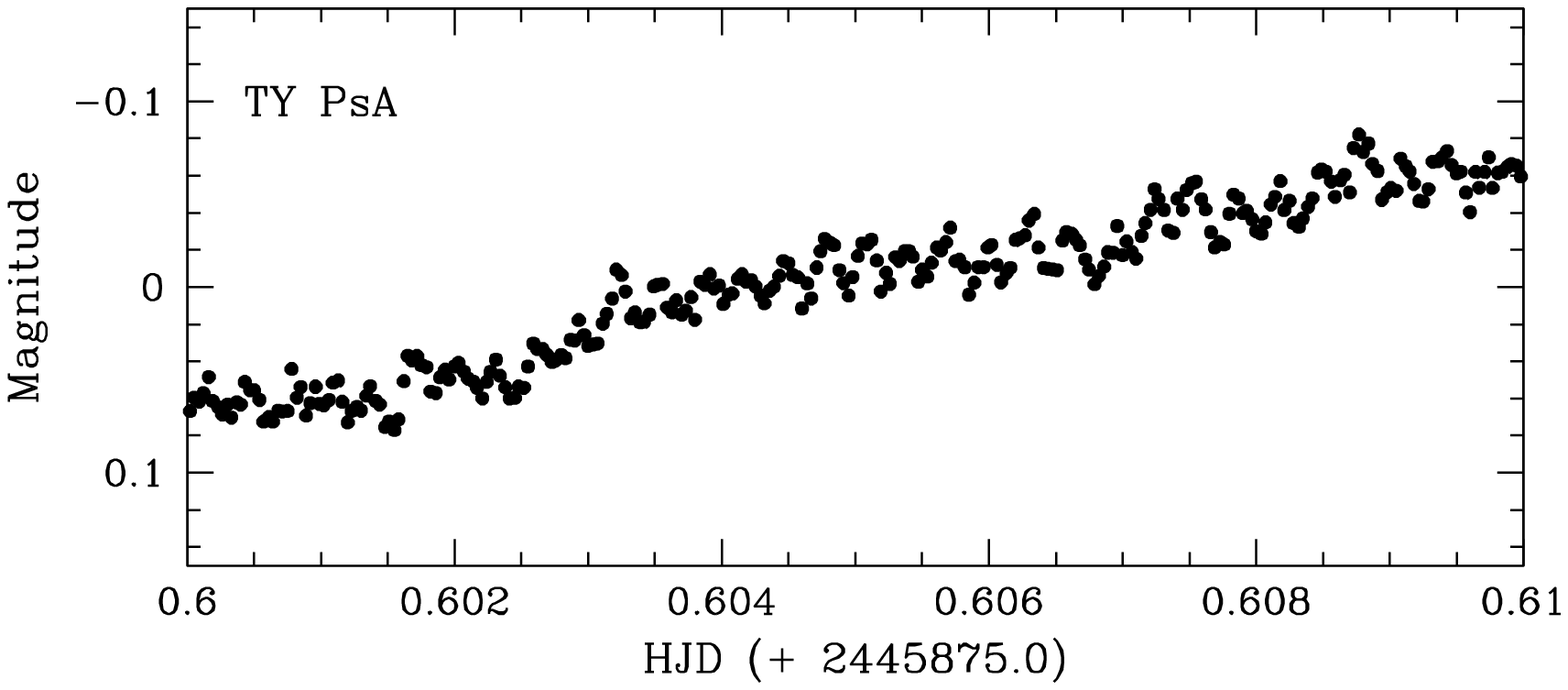}
\caption{The light curve of TY PsA. The DNOs are directly visible in the light curve. Individual points are 3 s integrations.
Adapted from Warner, O'Donoghue \& Wargau (1989).}
\label{fig1}
\end{figure}

   In the early searches for further examples of the DQ Her 
phenomenon (see Warner (1988) for a history of the application of 
high speed photometry to CVs) only one more nova, V533 Her, 
was added (Patterson 1979). However, the same time scale 
phenomenon, albeit of much lower stability and usually of very 
low amplitude, was found (Warner \& Robinson 1972) to be 
commonly present in dwarf novae during outburst and also in 
nova-like variables, i.e., in CVs of high rates of mass transfer 
($\dot{M}$). These modulations, with periods initially observed in the 
range 8 -- 40 s, became known as Dwarf Nova Oscillations 
(DNOs). In the optical they are usually of such low amplitude that 
they appear only in Fourier transforms, but occasionally they reach 
a few percent amplitude and can be seen directly in the light curve 
(Fig.~\ref{fig1}). In almost all cases 
they are highly sinusoidal in pulse profile. There is no apparent 
correlation between amplitude of DNOs and orbital inclination.

    The study of such rapid modulations in CV light curves used to 
necessitate the use of relatively inefficient photomultipliers, but the 
advent of high quantum efficiency CCDs, and the ability to 
window and bin their pixels, has made possible studies to much 
fainter levels, and in particular in crowded fields. An example of 
such a design, and its applications, is given in O'Donoghue (1995).

    Longer time scale modulations, of large amplitude but much 
lower coherence, were identified in CV light curves by Patterson, 
Robinson \& Nather (1977) and are commonly called Quasi-Periodic 
Oscillations (QPOs). As will be discussed below, these may have 
more than one time scale and physical origin.

    Recently, a new type of DNO has been recognised (Warner, 
Woudt \& Pretorius 2003; hereafter WWP), similar to the DNOs 
but having periods typically a factor of about four greater. These 
can co-exist with the DNOs and QPOs, and may at times be 
mistaken for ordinary DNOs.

   I will examine the general properties of each of these kinds of 
short time scale modulations. 
A general review of the literature on oscillations in 
CVs up to 1995 can be found in Warner (1995a). I omit from 
discussion here the rapid oscillations that are observed at $\sim$ 1 Hz 
in the accretion columns of strongly magnetic CVs (polars) -- e.g., 
Larsson (1992).

\section{DNOs}

    From the time of their discovery it was recognised that, unlike 
DQ Her, which has $Q = | {{dP}\over{dt}} |^{-1} \sim 10^{12}$, DNOs are low $Q$ 
modulations. On time scales of hours their periods can change by 
many seconds, but at other times are stable to milliseconds for 
many hours, giving $10^{3} < Q < 10^{7}$. Two time scales of variation are 
known. The first, on an outburst time scale, is a period-luminosity 
relationship, connecting DNO period to optical brightness, but not 
in a single-valued manner, thus producing ``banana loops'' in a 
diagram of $P_{DNO}$ versus $m_v$ (Patterson 1981). Although it was early 
hypothesised that $P_{DNO}$ would be found to be simply related to 
accretion luminosity (Bailey 1980), and that this would produce 
such loops because of the observed time delay between bolometric 
and optical luminosities (Hassall et al.~1983; Warner 1995b), it was 
long before this could be directly demonstrated as a monotonic 
relationship between $P_{DNO}$ and EUV luminosity (which is a good monitor
of bolometric accretion luminosity: Mauche 1996). It 
was even longer before simultaneous observations of DNOs in the 
optical and EUV showed identical, in phase, modulations (Mauche 
\& Robinson 2001).

   Although covering a smaller range of luminosities, the period-luminosity 
relationship has been demonstrated for nova-likes 
(Warner, O'Donoghue \& Allen 1985). It would be of interest to 
study any DNOs in nova-likes that go into states of low $\dot{M}$ (i.e., 
the VY Scl stars).

\begin{figure}
\plotone{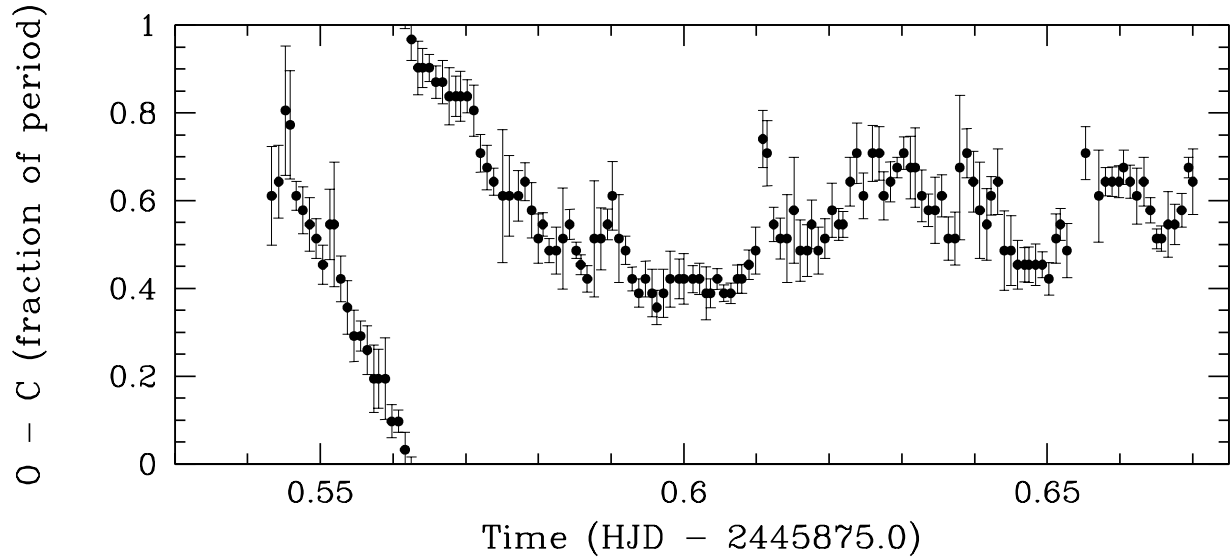}
\caption{The O--C diagram of TY PsA. From observations published in Warner, O'Donoghue \& Wargau (1989). The observed oscillations are compared with a sine wave of constant period 26.64 s.
At the start of the run the period was shorter than this, giving a nearly linear change of
phase (O--C), but around 0.58 d the period changed to average 26.64 s, but with short and
often sudden changes of phase (or, equivalent, period). The diagram ``wraps round'' when the 
phase exceeds one cycle.}
\label{fig2}
\end{figure}

    The second time scale of variation is very short: even when the 
DNOs are at their most coherent, changes are seen as sudden 
small jumps in period (typically $\sim$ 0.03 s, i.e. $\sim$ 0.1\%) that interrupt 
times of comparative stability lasting for thousands of seconds 
(Fig.~\ref{fig2}). There is evidence that DNOs are most stable 
at maximum of outburst but that their stability (i.e., coherence 
length) progressively decreases late in outburst (e.g. Hildebrand, 
Spillar \& Stiening 1981). The rarity with which any optical DNOs 
are seen at maximum light in VW Hyi, and their low amplitude  
($\sim$1 mmag) when observed (Woudt \& Warner 2002a: hereafter 
WW2a), could indicate that it is high stability that is rare, and that 
the usually employed Fourier transform techniques are inadequate 
for detecting drifting signals. Similarly, van der Woerd et al.~(1987)  
were successful in detecting DNOs on only two occasions during extensive soft X-Ray 
observations of 7 outbursts of VW Hyi, where the modulation 
is large ($\sim$ 15\%) when present.

     The night-to-night change from easy detectability to complete 
absence of DNOs in nova-likes (e.g. UX UMa, where DNOs are 
found in only about half of the optical and HST observations: 
Nather \& Robinson (1974); Knigge et al.~(1998)) also indicate 
cessation rather than reduced coherence. The changes of amplitude 
of DNOs over a few hours often include disappearance and 
reappearance.

   In addition to the abrupt changes in period, which are not 
accompanied by any noticeable change in system luminosity, there 
are short-lived phase excursions that appear to have no effect on 
the underlying trend in period. When best defined, the DNOs show 
the presence of an underlying clock with abrupt changes of period, 
on which the phase noise seems never to cause loss of knowledge 
of phase (Warner \& Brickhill 1978; Jones \& Watson 1992). 
However, the increasing frequency of the phase and period 
changes produces reduced coherence in late phases of outbursts. 
These short time scale variations are demonstrated in Fig.~\ref{fig2}. 

    When sufficient DNO coherence occurs in eclipsing dwarf 
novae or nova-likes, phase and amplitude variations can be 
measured during the eclipses. These show that, as in DQ Her, 
DNOs are caused by ``beams'' of radiation carried round by 
rotation of the primary, (Nather \& Robinson 1974; Patterson 1981; 
Petterson 1980). Confirmation comes from Keck time-resolved 
spectroscopy of V2051 Oph, which has a pulse phase in its emission 
lines that is wavelength dependent similar to that of DQ Her (Steeghs 
et al.~2001).

    An inventory of DNO observations\footnote{In many 
cases there has been only one observation of a DNO in a given star, but its persistence for an 
hour or more is sufficient evidence for its existence. The full range of period is in most cases not yet 
determined, and therefore the order of stars in Table 1 (where in the first part they are listed by increasing 
minimum observed period) is to some extent arbitrary. In a few cases of very low amplitude, short duration 
of appearance, or other reason, I have omitted published claims of DNO detections.} is given in Table 1. There 
are about 50 CVs in which standard DNOs have been detected, 
plus a few in which the nature of the DNOs is not yet determined. 
It is notable that none of the classic IPs appears in this list (an 
instructive exception is GK Per, see Section 4) -- despite much high 
speed photometry none of these has been reported to have 
modulations at periods shorter than their primary spin-related 
periods.

\section{LONGER PERIOD DNOs (lpDNOs)}

\begin{figure}
\plotone{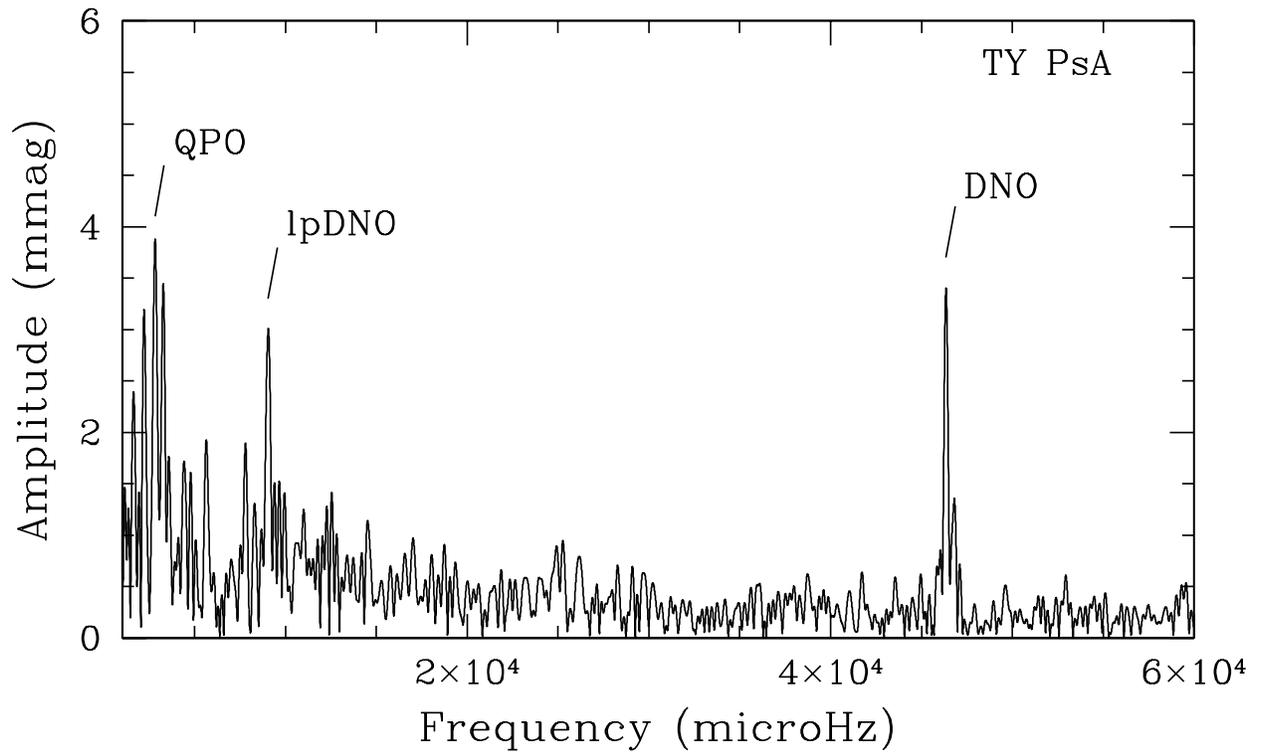}
\caption{The Fourier spectrum of TY PsA. The QPO, lpDNO and DNO are all simultaneously present. From unpublished
observations by M.L. Pretorius.}
\label{fig3}
\end{figure}

      It has recently been realised, from observing the literature as 
well as the sky, that there is another class of DNOs, showing 
coherence levels similar to DNOs but not following the same 
period-luminosity relationship -- in fact, their periods may be 
independent of $\dot{M}$ (WWP). We have called these `longer period 
DNOs' (lpDNOs); they have periods typically about four times 
greater than those of DNOs and can coexist with the latter. An 
example of this is shown in Fig.~\ref{fig3}. They are 
mentioned, but without explanation, in a number of early papers on 
DNOs -- specific examples are in VW Hyi, where a modulation 
with period averaging $\sim$ 88 s (in addition to $\sim$ 28 -- 36 s DNOs) was 
seen on seven nights of an outburst as the system decreased in 
brightness from magnitude 10.6 to 13.7 (Haefner, Schoembs \& 
Vogt 1979; examples of similar, but rare, periodicities in VW Hyi 
are given by WWP); SS Cyg, where $\sim$ 33 s modulation was present 
in addition to 10 s DNOs (Robinson \& Nather 1979; Patterson 
1981); HT Cas and AH Her, where $\sim$ 100 s modulations were seen 
(the former even in quiescence) but where normal DNOs have 
much shorter periods (Patterson 1981).

    lpDNOs have now been observed in about 17 CVs (Table 1); 
these include some reclassifications, and there may be other cases 
where DNOs have mistaken identity. No studies of phase shifts 
during eclipse have yet been made, but the existence of double 
lpDNOs (Section 4.1) indicates that a rotating beam model is 
applicable and therefore phase shifts should occur.

\begin{figure}
\plotone{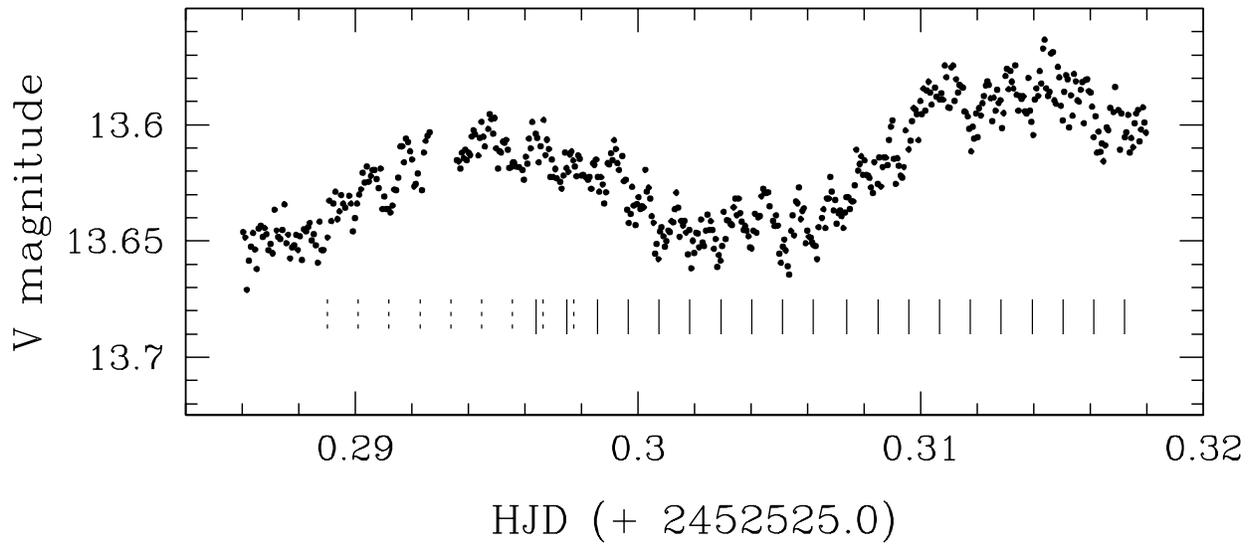}
\caption{The light curve of EC\,2117-54 (the first 45 minutes of run S6554). The lpDNO modulation
at 94.21 s is clearly visible in the light curve. The lpDNO minima are marked by vertical bars. There is
a phase shift around HJD\,2452525.297. From Warner, Woudt \& Pretorius (2003).}
\label{fig4}
\end{figure}

   As with DNOs, the lpDNOs are not always present; on average 
they have slightly larger amplitudes than DNOs and are more often 
seen directly in the light curve -- see Fig.~\ref{fig4}. 
For this reason some previously observed lpDNOs were classified 
as QPOs.

    Comparison of the amplitude and phase changes of DNOs and 
lpDNOs when both are simultaneously present shows that they 
generally behave independently (WWP).

    Table 1 includes four AM CVn stars -- helium-transferring CVs 
with ultra-short orbital periods. The detection of the various types 
of rapid modulation in AM CVn stars is made difficult by the 
presence of strong harmonics to the main orbital/superhump 
brightness variations. (These harmonics are not independent 
oscillations in the system -- they appear in Fourier transforms 
merely from the highly nonsinusoidal shapes of the principal 
modulations.)

     Nevertheless, V803 Cen has long been known to possess a 
modulation at $\sim$ 178 s (O'Donoghue, Menzies \& Hill 1987) of 
unknown origin. It is seen when V803 Cen has a high $\dot{M}$, but 
not during low states, its period appears independent of luminosity, 
and there are phase variations on time scales of hours, similar to 
those of DNOs (WWP). These are the characteristics of lpDNOs, 
rather than, e.g. possible pulsations of the primary. No ordinary 
DNOs have yet been detected in V803 Cen, but its compatriot CR 
Boo has shown the full set of DNO, lpDNO and QPO oscillations 
(WWP) and the type star AM CVn has shown both ordinary DNOs 
and two periodicities in the QPO range.
   
\section{QUASI-PERIODIC OSCILLATIONS}

\begin{figure}
\plotone{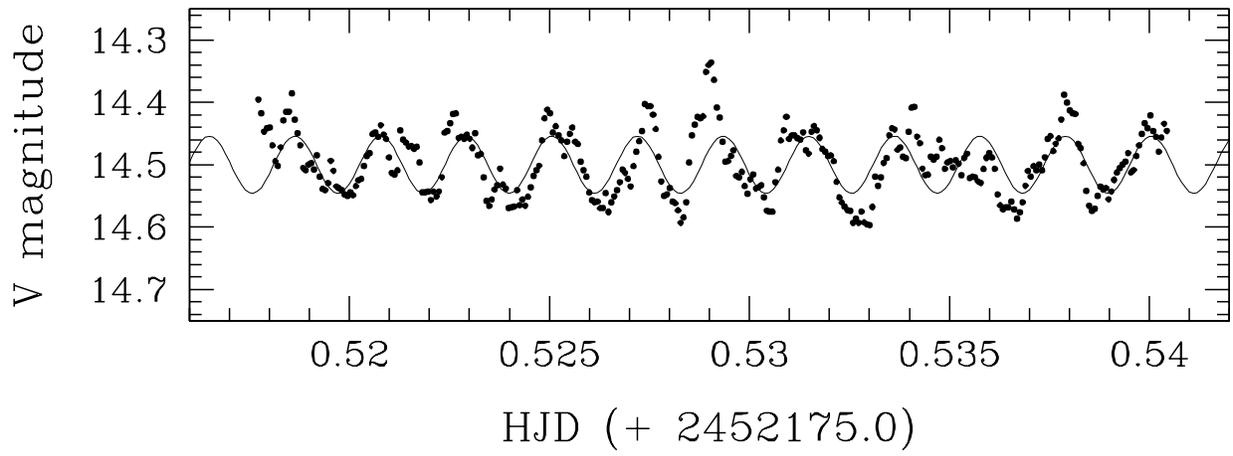}
\caption{The light curve of WX Hyi, showing the 185-s QPO clearly. Superimposed is the result
from the non-linear sinusoidal least-squares fit. From Warner, Woudt \& Pretorius (2003).}
\label{fig5}
\end{figure}

     Quasi-periodic modulations have orders of magnitude less 
coherence than DNOs. Because of this they are difficult to 
recognise in Fourier transforms -- their power is spread over a wide 
band of frequency. They typically have $Q \sim 5 - 20$. An example of 
an unusually coherent train of 11 cycles of a QPO is shown in 
Fig.~\ref{fig5}; a Fourier transform of this light curve 
shows that 19.4 s DNOs with a mean amplitude of 3.8 mmag were 
also present.

     In principle there may be more than one kind of variation, e.g. 
almost constant period but with frequent large phase changes, with 
perhaps a clear growth and decay in amplitude between the phase 
jumps, or constantly varying period and amplitude over a limited 
range. There has as yet been no study or classification of such 
modulations; their analysis is hampered by the general stochastic 
flickering, arising from accretion processes, that is usually present 
in CVs (though here again some reclassification may be necessary -- some 
of the flickering and flaring commented on in the CV 
literature has a QPO look about it). More refined definitions and 
analyses will probably reveal that QPOs are more common than 
hitherto realised -- at present we accept only the QPOs in a light 
curve that are obvious to the eye.

    A glance down the column listing QPO periods in Table 1 will 
indicate the evident existence of two time scales of QPO. In a few 
cases (e.g. WX Hyi, VZ Pyx, OY Car, perhaps V2051 Oph) two 
quite different periods of QPOs have been observed. For those 
systems where DNOs and/or lpDNOs are seen there are often 
QPOs at $\sim$ 15 times the DNO period or, equivalently, $\sim$ 4 times the 
lpDNO period. Apart from these there is a large number of 
observations of quasi-periodic modulations with periods in the 
range 1000 -- 3000 s (note that GK Per is an exception here -- its 
5000 s QPO is $\sim$ 15 times its optical DNO period). I have to 
distinguish between these two kinds of QPO here, but am reluctant 
to introduce yet more nomenclature, and so will simply discuss 
separately those that are related to DNO/lpDNO periods and those 
that are not.

\subsection{DNO-related QPOs}

     A valuable clue to the nature of these QPOs is given by the 
existence of ``double DNOs''. Pairs of DNOs with small 
separations in period were found in V3885 Sgr (29.08 and 30.15 s: 
Hesser, Lasker \& Osmer 1974), V2051 Oph (29.77 and 28.06 s: 
Steeghs et al.~2001), OY Car (17.94 and 18.16 s: Marsh \& Horne 
1998), V436 Cen (19.45 and 20.20 s: WWP), EC2117-54 (many 
examples, typically 22.10 and 23.27 s: WWP), CN Ori (11.23 and 
12.10 s: WWP) and VW Hyi (see below: WW2a) -- and there is the 
case of WZ Sge, which shows stable modulations at 27.87 and/or 
28.95 s even in quiescence (Table 1), which I will discuss in 
Section 9.

\begin{figure}
\plotone{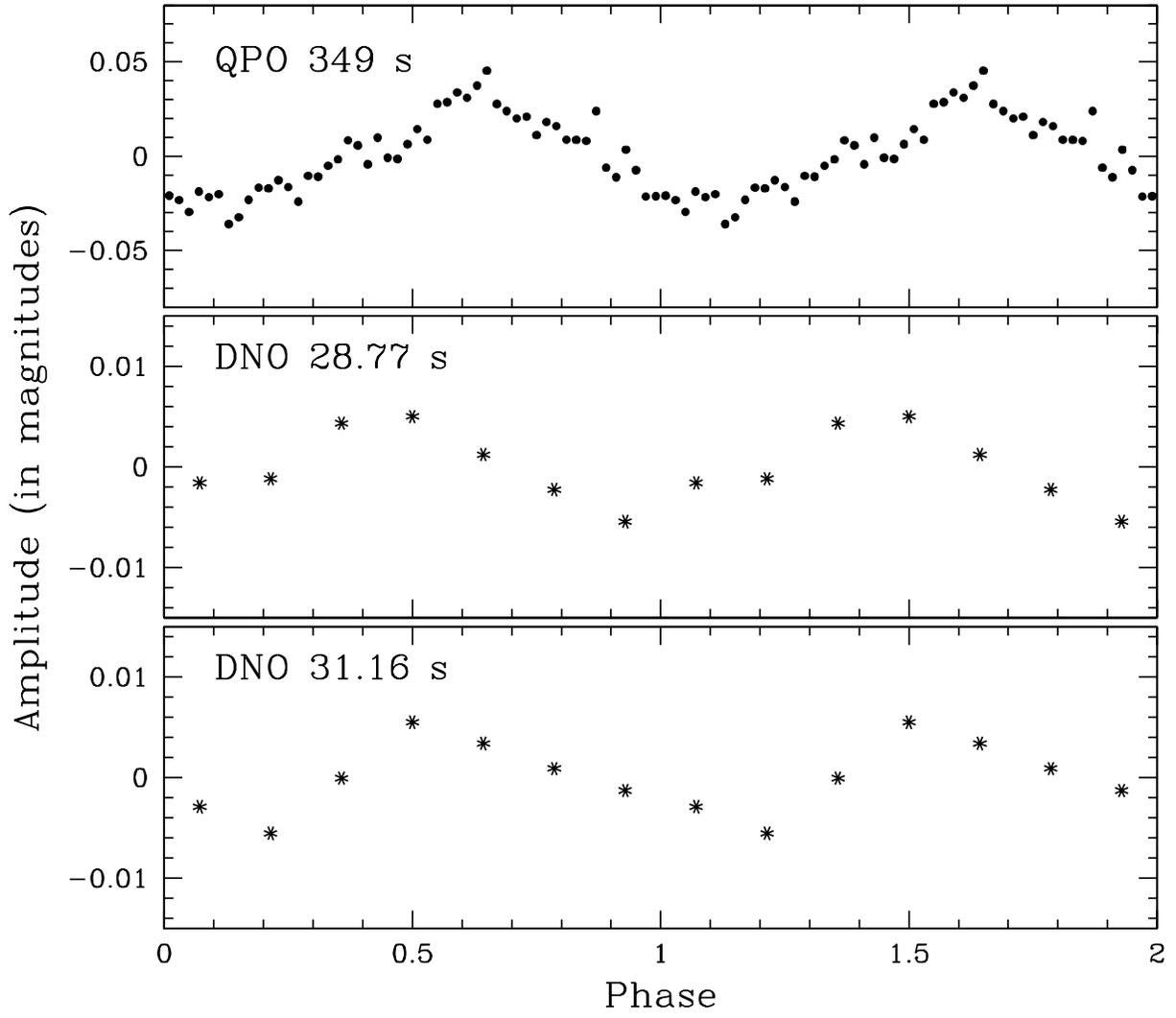}
\caption{Averaged profiles of the DNOs and QPOs present simultaneously in VW Hyi (corresponding to 
the Fourier spectrum in Fig.~\ref{fig12}). From Woudt \& Warner (2002a).}
\label{fig6}
\end{figure}

   In many of these cases the beat period between the double DNOs 
corresponds to the period of a QPO present at the same time -- 
including WZ Sge (Warner \& Woudt 2002b: hereafter WW2b). In 
Fig.~\ref{fig6} the mean pulse profiles of two DNO 
components and the accompanying QPO are illustrated. An 
example of a double lpDNO, split at the QPO frequency, has been 
found in VW Hyi (WWP). 

\begin{figure}
\plotone{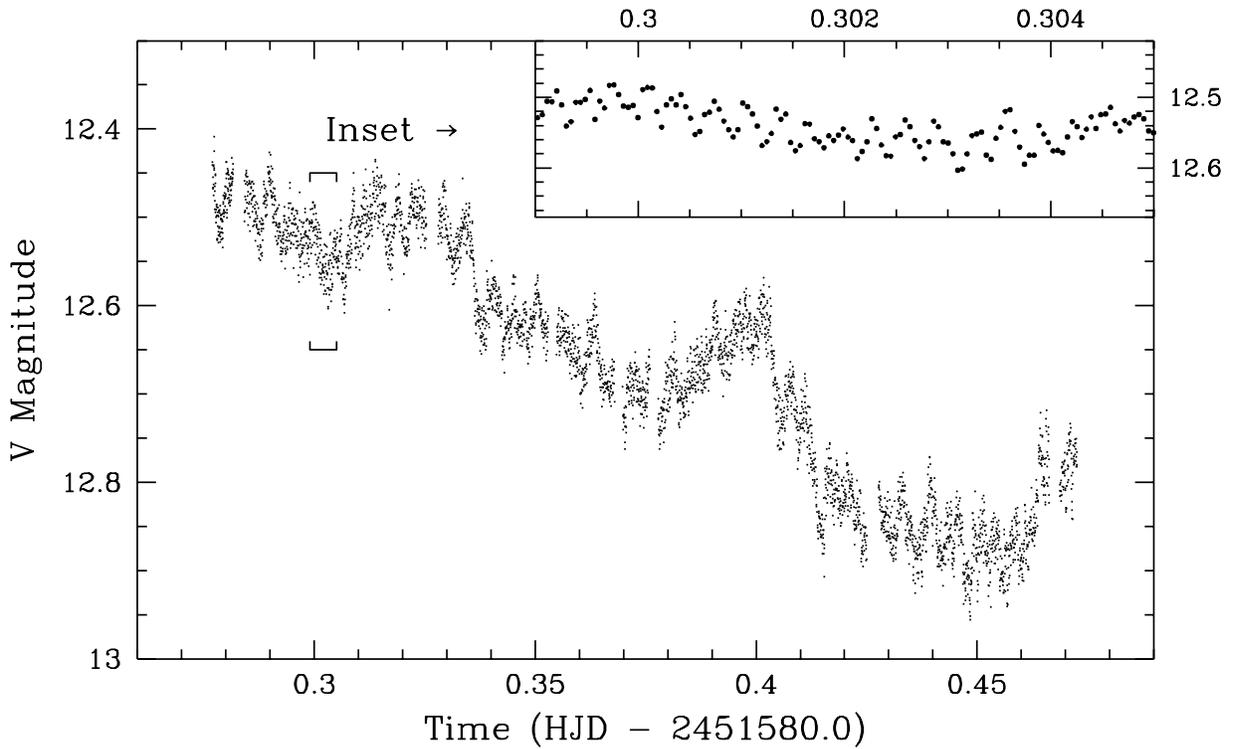}
\caption{The light curve of VW Hyi on 2000 Feb 5, taken during the late decay 
phase of this dwarf nova outburst. The inset is an amplified view of a small part, 
showing the DNOs. The large humps are at the orbital period (107 min). QPOs are
present with a range of $\sim$ 0.1 mag and time scale $\sim$ 0.006 d.
From Woudt \& Warner (2002a).}
\label{fig7}
\end{figure}

\begin{figure}
\plotone{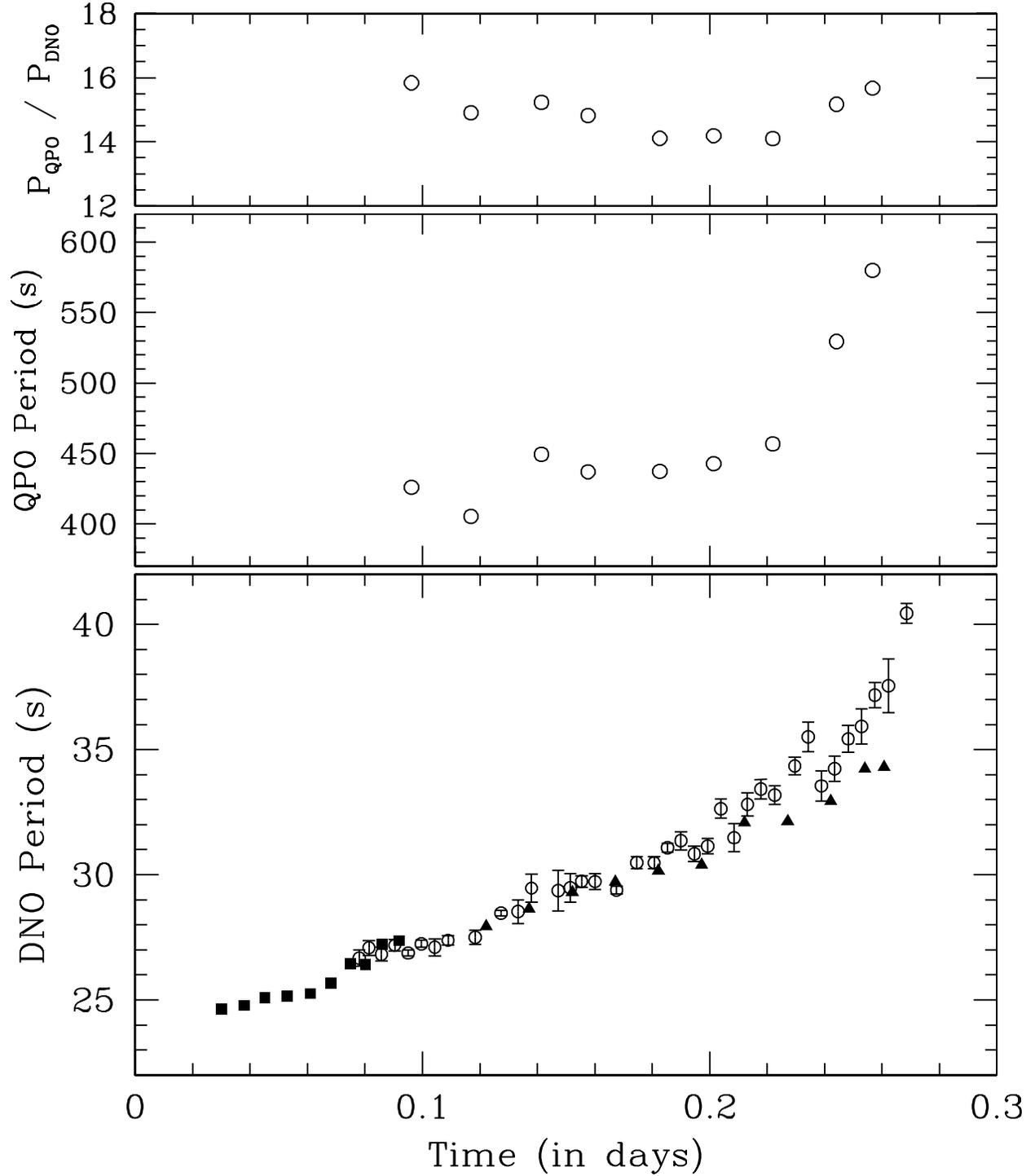}
\caption{Variations with time of the DNO and QPO periods in the normal outburst of 2000 February (circles with error bars).
DNOs are added for the superoutburst of 1972 December (triangles) and the normal outburst of 2001 February (squares). The topmost
panel shows the ratio of the periods in the 2000 February run. From Woudt \& Warner (2002a).}
\label{fig8}
\end{figure}

   Another clear example of an interrelationship between DNOs 
and QPOs is given by the light curve of VW Hyi for 5 Feb 2000 
(Fig.~\ref{fig7}), obtained at the end of outburst when 
large amplitude DNOs and QPOs typically occur in this star. The 
fall in brightness by a factor of about two during the observation 
caused steady increases in period for both the DNOs and the QPOs -- the 
first time such an evolution of QPO period had been seen. 
The ratio $P_{QPO}$/$P_{DNO}$ remained at about 15 during this evolution 
(Fig.~\ref{fig8}). This is generalised in Section 10. 
     
\subsection{Other QPOs}

     Even among the remaining QPOs there may be two or more 
distinct types. Many of the CVs listed in Table 1 with periods  
$\sim$ 1000 s are suspected by Patterson et al.~(2002b) to be generated by 
IP structures -- as in the case of GK Per where the underlying 351 s 
rotation of the primary is seen as $\sim$ 380 s modulation caused by 
reprocessing of the rotating beam off a varying period QPO source. 
The periods are characteristic of the ``canonical'' IPs, which have 
rotation periods $\sim$ 15 min and binary periods $\sim$ 4 h. Included 
among these systems are LS Peg and V795 Her, both of which 
show polarization modulated at the optical periods (Rodriquez-Gil 
et al.~2001, 2002), but which have not yet been proved to have the 
high stability required for definite IP status.

   For these systems, therefore, the evidence is accumulating that 
their QPOs are another manifestation of magnetic primaries -- and 
are occurring in IPs for which large amplitude coherent optical and 
X-Ray modulations are in some way suppressed -- perhaps by high 
$\dot{M}$. These will therefore probably add directly to the tally of IPs.

   But, anticipating the model developed in Section 6, a CV cannot 
have DNOs (and/or lpDNOs) and be an IP -- the magnetic field is 
either strong enough to anchor the exterior of the white dwarf
to the interior, or it is 
not -- and this exclusivity is indeed demonstrated by the absence of 
any observed DNOs in known IPs. So we are left with at least a 
few CVs for which the QPOs cannot have a direct connection with 
rotation of the primary. In Table 1 these are the systems WX Hyi, 
VZ Pyx, OY Car and V2051 Oph, already mentioned above, and 
there may be others in which only QPOs have so far been detected. 
The time scales of some of these modulations are comparable with the 
rotation periods at the outer edges of a high $\dot{M}$ accretion discs 
(which are $\sim$ 0.20 $P_{orb}$ -- equation 8.6 of Warner 1995a), but no 
definite models accompany this speculation. 

   Another possible source of brightness variation is modulation of 
$\dot{M}$ from the secondary, caused by non-radial oscillations of the 
secondary in a manner similar to that of the $\sim$ 5 min oscillations of 
the Sun. Small amplitude variations in radius of the secondary 
would be amplified in $\dot{M}$ by the great sensitivity of Roche lobe 
overflow.

\section{X-RAY OBSERVATIONS}

     CVs, even nominally `non magnetic' ones, are usually 
detectable in X-Rays: the low $\dot{M}$ systems have hard X-Ray 
emission and the high $\dot{M}$, optically thick, systems have soft X-Rays. 
A transition from hard to soft and back is seen during a 
dwarf nova outburst. 

      Table 2 lists X-Ray observations of DNOs and QPOs in CVs -- 
it includes only the lower $Q$ modulations, i.e., it omits IPs. The 
best-studied systems are SS Cyg, U Gem and VW Hyi, which are 
close enough to the Sun for soft X-Rays to be received almost 
unabsorbed by interstellar gas and are thus very strong sources 
during outbursts. There appears to be a mixture of DNOs, lpDNOs 
and QPOs, but VW Hyi and U Gem in particular have full sets 
(note that in the case of VW Hyi the DNOs were observed at 
maximum but the QPOs were at the very end of outburst, so the 
ratio $P_{QPO}$/$P_{DNO}$ is not close to the value 15 seen in the optical). 
Where there is an overlap, it is evident that individual stars show 
the same modulation time scales in optical and X-Ray.

\begin{figure}
\plotone{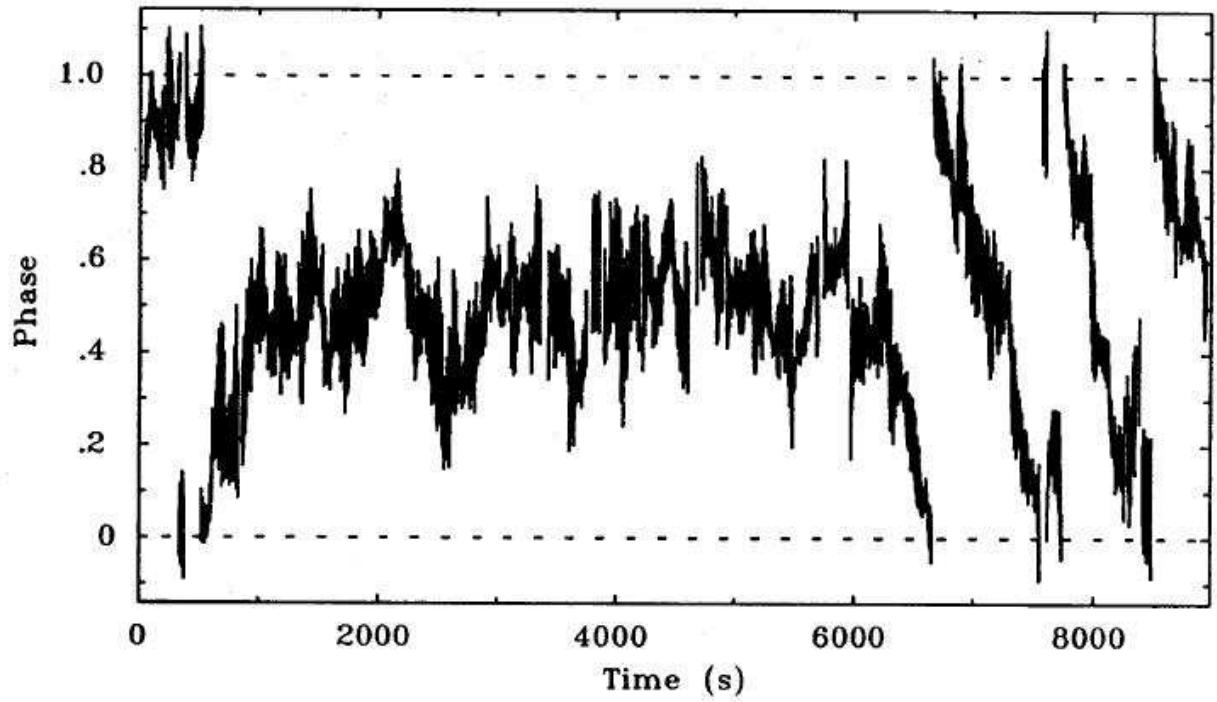}
\caption{Phase behaviour of DNOs in SS Cyg in X-Rays (adapted from Jones \& Watson 1992).
This is the same kind of diagram as Fig.~\ref{fig2}, displaying O--C phase relative 
to a constant period.}
\label{fig9}
\end{figure}

   The phase behaviour of DNOs is the same in X-Rays as in the 
optical. The most detailed study is that of Jones \& Watson (1992); 
one of their results is shown in Fig.~\ref{fig9}, 
which should be compared with Fig.~\ref{fig2}. The amplitude of 
modulation in this case averaged about 35\% but at times reached 
nearly 100\%. Such large amplitudes are commonly seen in soft X-
Ray DNOs.

    Table 2 shows that ordinary DNOs seen during dwarf nova 
outbursts occur in soft X-Rays but not hard X-Rays. This is most 
notable in the comprehensive study by Wheatley, Mauche \& 
Mattei (2003: hereafter WMM) of an outburst in SS Cyg. The few 
observed hard X-Ray DNOs have all occurred at quiescence. 
WMM conclude that the source of the DNOs must lie in the 
optically thick boundary layer at the surface of the primary. 
Although `residual' hard X-Rays are observed during outburst, 
these are not modulated as DNOs -- but this is in accordance with 
the interpretation of these X-Rays as being coronal in origin 
(WMM).

    On the other hand, modulations at QPO periods are seen in hard 
X-Rays, most strongly in VW Hyi just as the hard X-Rays are turning 
on at the end of an outburst (Wheatley et al.~1996) and in SS Cyg 
during the turn-off phase of hard X-Rays after the start of outburst 
and again when they turn on at the end of outburst (WMM). These 
phases correspond to the transitions between optically thick and 
optically thin discs, which themselves are governed by $\dot{M}$ in the 
inner disc. The QPOs themselves are not accompanied by any 
hardness variations, thus eliminating temperature variations or 
photoelectric absorption variations as the source of the oscillations 
-- but it does allow quasi-periodic occultation by a source that is 
completely opaque at X-ray energies (WMM).

\section{STATISTICAL VERSUS PHYSICAL MODELS}

     A number of statistical analyses of DNO light curves have been 
undertaken. Early models included a damped harmonic oscillator 
excited by white noise (Robinson \& Nather 1979; Hildebrand, 
Spillar \& Stiening 1981), and a sinusoidal oscillator exercising a 
random walk in phase (Horne \& Gomer 1980; Cordova et al.~1980, 
1984). One problem with such approaches is that although they 
may deliver quantitative values of statistical parameters these do 
not necessarily lead to any physical insight. A worse problem 
arises when it is seen that, for the best quality data, such analyses 
are inappropriate -- e.g., Jones \& Watson (1992) show that the 
random walk model does not represent the behaviour of the soft X-Ray 
DNOs of SS Cyg. But that should be no surprise given the 
systematic behaviour seen in the phase diagrams (Figs.~\ref{fig2} and \ref{fig9}).

    A more profitable approach, therefore, is to start with the 
observed properties and try to envisage a model that can reproduce 
them. In presenting what I think is a viable model I am influenced 
by the following facts:

\begin{itemize}
\item{At least 10\% of single white dwarfs have magnetic fields 
$> 2 \times 10^6$ G, and the limited studies with sensitivities down 
to $\sim 3 \times 10^4$ G suggest that the total may be significantly higher (Liebert, 
Bergeron \& Holberg 2003). At least 25\% of white dwarfs in CVs 
have detectable or directly inferable fields (i.e. they are polars or 
IPs), despite the fact that the current lower limit for detection in 
these systems is $\sim 7 \times 10^6$ G (Wickramasinghe \& Ferrario 2000). 
But here there may be a selection effect operating: most strongly 
magnetic CVs have been found from X-Ray surveys, which probe 
to greater distances than extant optical surveys (though the Sloan 
Digital Sky Survey may eventually remove this bias). On the other 
hand, at least some of the high $\dot{M}$ IPs avoided detection until 
recently (e.g. the SW Sex systems LS Peg and V795 Her). It has 
been suggested that in the common envelope phase magnetic fields 
are generated and amplified within the differentially rotating 
envelope (Regos \& Tout 1995), so it would not be surprising to 
find that CV primaries are systematically more magnetic than 
isolated white dwarfs.}
\item{Given this we may suspect that, not only are there many 
primaries in CVs with fields $< 7 \times 10^6$ G, they may even constitute 
a dominant fraction of the apparently ``non magnetic'' majority. 
The DNO model that I favour is based on this expected extension 
of the distribution of magnetic fields of primaries to values lower 
than those in IPs.}
\end{itemize}

\section{A MAGNETIC MODEL FOR DNOs and lpDNOs}

     The observed period-luminosity relationship for DNOs 
prompted Paczynski (1978) to suggest that magnetically 
channelled accretion was responsible, but onto an equatorial belt 
that slips on the surface of the primary, rather than onto the body 
of the primary itself. This was developed by Warner (1995b) and 
more recently in WW2b. Direct evidence for long-lived, rapidly 
rotating equatorial belts on the primaries of dwarf novae after 
outburst has come from HST observations  (Sion et al.~1996; 
G\"ansicke \& Beuermann 1996; Cheng et al.~1997; Szkody et al.~1998; 
Sion \& Urban 2002), where spectra are found to be 
composite, comprising a white dwarf in relatively slow rotation 
plus a hot belt rotating at a rate comparable to the Keplerian 
velocity at the surface of the primary.

     I will describe the model briefly, and show how it relates to the 
principal observed properties of DNOs, lpDNOs and QPOs.

    An accretion torque applied to the surface of a star can only be 
communicated to the bulk of the star if the viscosity of the interior 
of the star is large enough. Degenerate material is notoriously 
slippery, so a white dwarf behaves like a solid body only if it is 
permeated by a strong enough magnetic field (Durisen 1973). 
~Strong enough' turns out to be $B > 1 \times 10^5$ G (Katz 1975, WW2b). 
For accretion to be magnetically controlled near the surface of the 
primary $B$ must also be strong enough -- depending on $\dot{M}$ and 
the geometry of the field. For $\dot{M} \sim 5 \times 10^{17}$ g s$^{-1}$, probably the 
maximum reached in dwarf novae in outburst, the requirement (for 
a multipole field) is $B > 2 \times 10^4$ G (equation 16 of WW2b). 

     There is therefore a window of opportunity -- even high $\dot{M}$ 
accretion onto low field primaries can be magnetically controlled 
by whatever field the freely slipping equatorial belt has. The field 
within the belt itself will be enhanced during accretion, by 
differential shearing of the field lines, which aids magnetic 
channelling; this may also explain why dwarf novae in quiescence 
in general do not show DNOs. CVs with very low intrinsic fields 
should never show DNOs: several dwarf novae and nova-likes 
have been extensively observed with high time resolution without 
any sign of DNOs (Patterson (1981) and Warner (1995a) list 
examples) -- these constitute the genuine non-magnetic CVs.

    The model is therefore similar to that for an IP, but is of low $Q$: 
the mass of the belt accreted during a dwarf nova outburst 
(obtained by measuring the total energy radiated, $\sim 10^{39}-10^{40}$ erg, and 
ascribing it to released gravitational energy) is $\sim 10^{22}$ g (cf. $10^{33}$ g 
for the whole primary) and can thus be easily tugged around by 
magnetic coupling to the accretion disc. It spins up as $\dot{M}$ 
increases on the rising branch of outburst, which compresses the 
magnetosphere of the primary and decreases the period at the inner 
radius of the disc. Because $\dot{M}$ varies with time during an 
outburst, no equilibrium is reached. As the inner radius of the disc 
steadily decreases, the minimum energy state for accretion is 
constantly being sought, so field lines reconnect between the 
equatorial belt (which has differential rotation in latitude as well as 
in depth) and the disc, resulting in accretion along field lines 
connected to zones of slightly different periods of rotation. This 
search for equilibrium causes the sudden jumps of DNO period as 
matter is transferred from one magnetic channel to another; they 
are least frequent at maximum luminosity when $\dot{M}$ passes 
through its turning point and hence is most constant.

   After maximum luminosity, $\dot{M}$ decreases steadily, and at 
some point the inner edge of the disc retreats so rapidly that the 
belt is not able to slow its rotation period rapidly enough to 
maintain near equilibrium. Then gas attaching itself to the field 
lines anchored in the belt is centrifuged outwards, stopping most of 
the accretion onto the primary and extracting angular momentum 
from the belt. 

\begin{figure}
\plotone{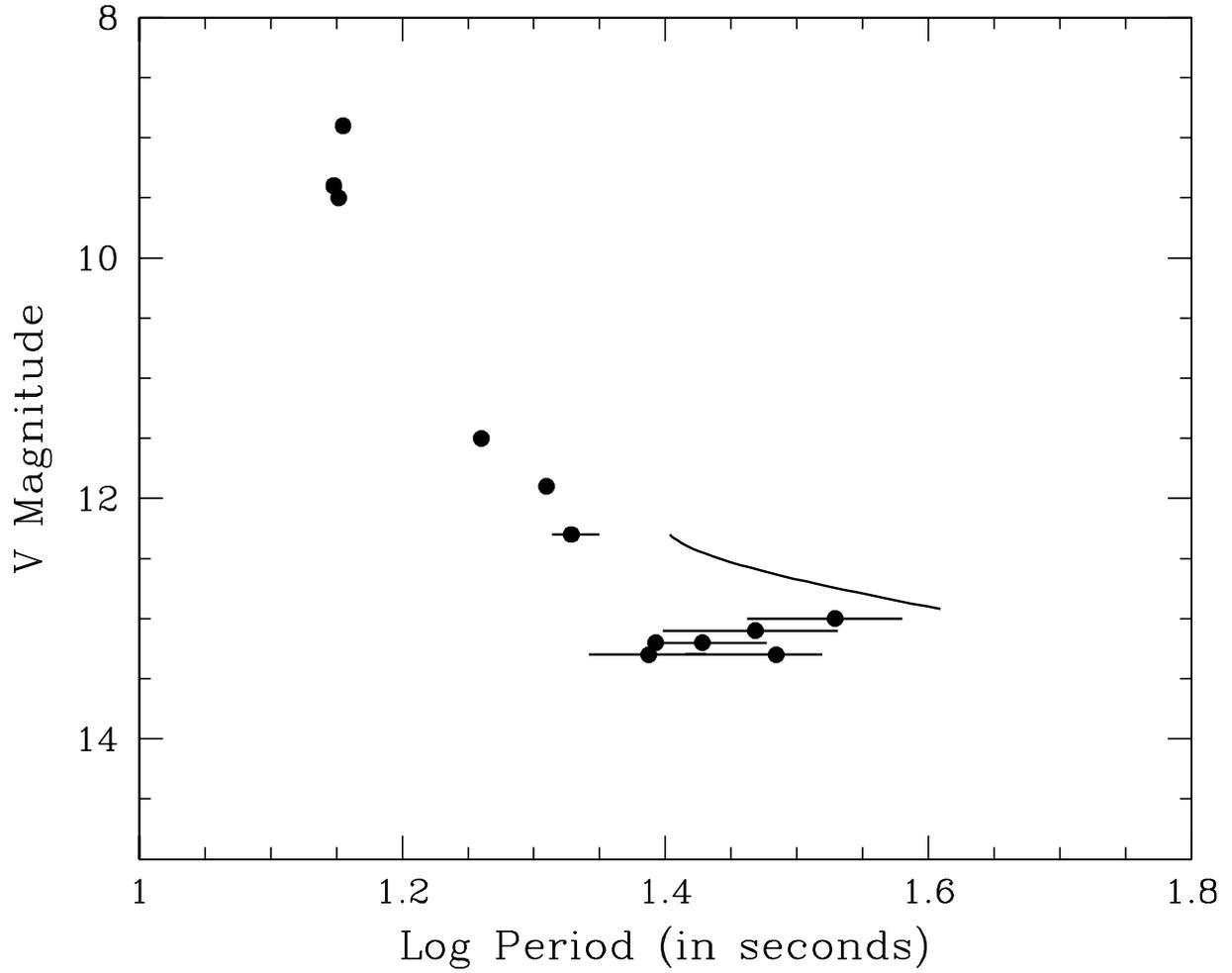}
\caption{DNO periods as a function of the V magnitude of VW Hyi. The curved continuous line corresponds to the DNO evolution
seen in Fig.~\ref{fig8}. The horizontal bars show the range of DNO periods in each of the runs illustrated.
From Woudt \& Warner (2002a).}
\label{fig10}
\end{figure}

\begin{figure}
\plotone{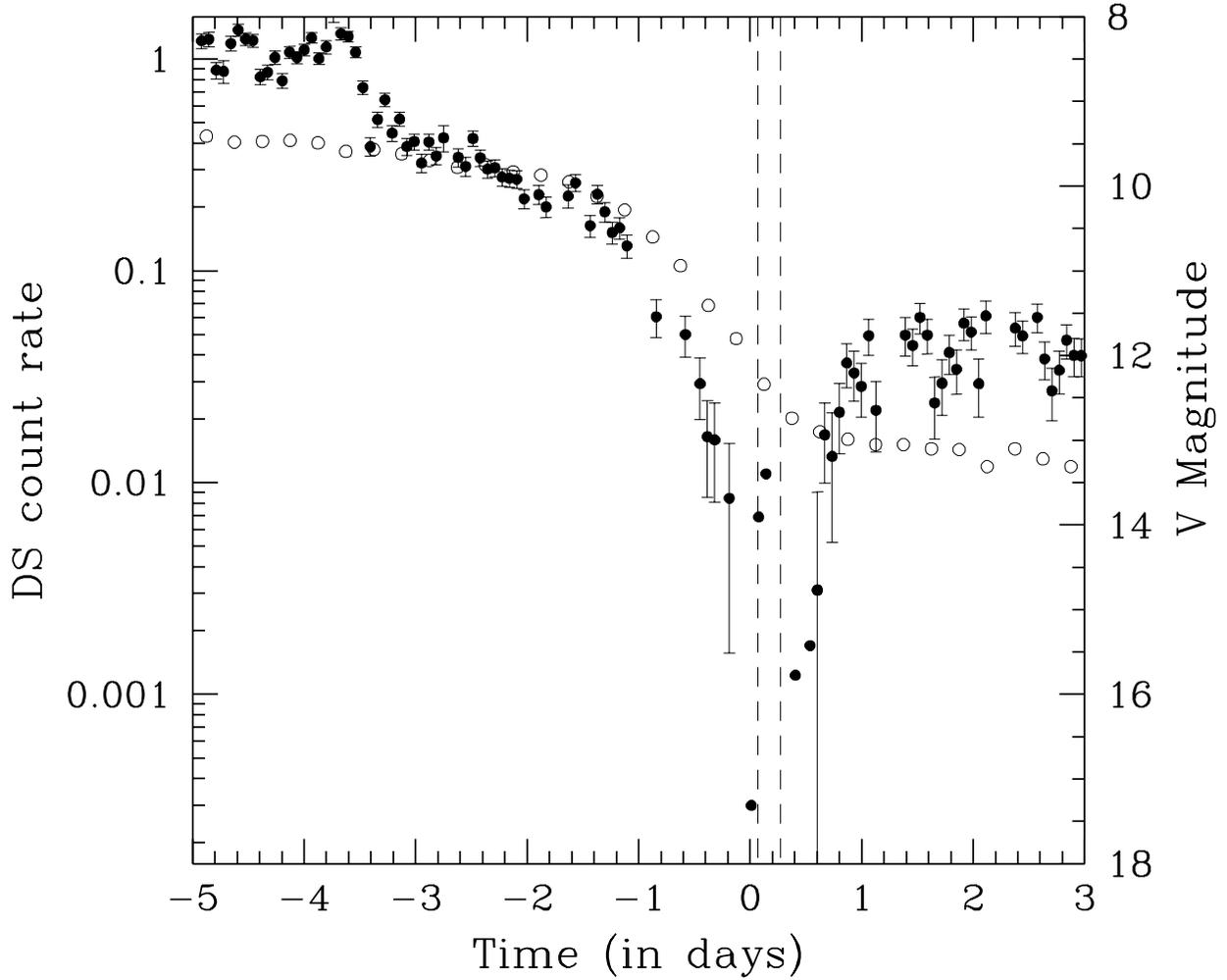}
\caption{Comparison of EUV flux (DS count: filled circles) at the end of a superoutburst in VW Hyi with the average optical
light curve (open circles). The vertical dashed lines show the range over which the DNOs in Fig.~\ref{fig8} were observed.
EUV data are from C.W. Mauche (personal communication), optical data are from the Royal Astronomical Society of New Zealand
(provided by F. Bateson).}
\label{fig11}
\end{figure}

    Evidence for such `propellering' is seen in VW Hyi. The 
variation in DNO period as a function of brightness on the 
descending branch of outburst is shown in Fig.~\ref{fig10}. 
The general correlation between period and brightness for 
$8.3 < V < 12.3$ is an example of the ubiquitous period-luminosity 
relationship referred to above. But that is succeeded by a phase of 
extremely rapid increase in period, during which the DNO period 
doubles in about 6 hours. This phase is concomitant with cessation 
of EUV flux (Fig.~\ref{fig11}). As EUV flux is a 
competent monitor of $\dot{M}$ onto the primary, it can be seen that 
accretion is reduced to a trickle exactly at the time that the deduced 
rapid deceleration of the equatorial belt is occurring.

    What about lpDNOs? Their independence from accretion 
luminosity puts them in a different category from ordinary DNOs, 
but in other aspects they behave like magnetically channelled 
accretion, including the existence of double lpDNOs split at the 
QPO frequency (Section 4.1). A clue to the cause of their longer 
periods comes from observed rotationally broadened spectra of 
some CV primaries. Taking the observed projected rotational 
velocities (Sion 1999) and adopting measured or estimated masses 
and inclinations (details are given in WW2b and WWP) we find 
the following rotation periods (all with at least 20\% uncertainty): 
VW Hyi: 140 s; SS Cyg: 63 s; OY Car: $>$260 s. These rotation 
periods are roughly twice the observed lpDNO periods, which 
would occur for accretion onto two regions of the primary. I 
suggest, therefore, that lpDNOs are literally connected to the 
primary's rotation, which will be differential in latitude as material 
accreted at the equator spreads out across the surface. This 
explains why, unlike DNOs, lpDNO periods are not correlated 
with $\dot{M}$ and they behave independently of DNOs.

    If this interpretation is correct then the lpDNOs allow an 
extension of knowledge of rotation periods -- from the IP region to 
primaries of lower field strength. The definite lpDNOs in Table 1 
cover the range 33 -- 177 s. If we exclude AE Aqr, then the 
presently known IPs have rotation periods that range from 142 s of 
DQ Her to 4021 s for EX Hya. The reason for omitting AE Aqr is 
because the field of $\sim 3 \times 10^5$ G deduced for it (Wynn, King \& 
Horne 1997; Choi \& Yi 2000) is at the bottom end of the range of 
IP fields and may not be quite large enough for the primary to 
behave as a solid body. This in turn calls into question the high 
spin-down power that is deduced from the observed large $dP_{rot}/dt$ 
in AE Aqr (de Jager et al.~1994): indeed, the rapid spin-down may 
be simply because only the inertia of the outer layers of the 
primary is coupled to the retarding torque (which, because of the 
high rotation rate, is largely due to propellering).

    The existence of two clear groups -- rapidly rotating primaries 
with signatures of weak fields, and slower rotating IPs with quite 
strong fields -- constitutes a correlation between magnetic field 
strength and rotation period of the primary. This is related to the 
fact that the stronger the field, the larger is the radius of the 
magnetosphere, and the longer is the IP equilibrium period. But 
$\dot{M}$ also plays a role -- the larger the $\dot{M}$, the smaller the 
equilibrium radius of the magnetosphere. And from the observed 
time scales ${{P_{rot}}\over{| dP_{rot}/dt |}} \sim 10^6$ y in IPs we see that it is the average 
$\dot{M}$ over the past millions of years that determines the current 
value of Prot. Such a time scale probably includes several nova 
explosions, and may include long periods of very low $\dot{M}$ 
(hibernation -- see Section 9.4.3 of Warner 1995a), so the apparent 
correlation between field strength and spin period means either that 
the presently weak field CVs have preferentially had historic high 
$\dot{M}$, or that all CVs experience roughly similar average $\dot{M}$ and 
that their spin periods are solely a result of their weaker fields. The 
latter scenario seems more probable, and, indeed, if measurements 
of $B$ can be made, offers a way to estimate long-term average 
values of $\dot{M}$.

     I have suggested that DNOs and lpDNOs are the signatures of 
magnetically channelled accretion onto regions that are rotating 
respectively with maximum angular velocities close to the 
equatorial Keplerian velocity and at roughly one eighth (for two-pole
accretion) of that 
velocity. According to Kippenhahn \& Thomas (1978), the large 
specific angular momentum of material accreted at the equator 
onto a non-magnetic primary results in very slow mixing to higher 
latitudes: the time scale is much greater than the time between 
nova eruptions (but note that this computation includes only 
mixing via the Richardson instability; coupling of parts of the 
surface by closed loop magnetic fields will also distribute angular 
momentum). They find that the latitudinal width of the equatorial 
belt, caused by mixing and defined as that latitude where the 
velocity falls to half of the equatorial velocity, is $\sim 20^{\circ}$ (beyond 
which it falls very rapidly). They also find that $10^4$ y of accretion at 
$10^{-9}$ M$_{\odot}$ y$^{-1}$ results, after mixing with surface material, in a belt 
with an angular velocity 0.22 of the Keplerian velocity in the 
equatorial plane -- and therefore $\sim$ 0.11 at latitude 20$^{\circ}$. In the 
suggested model for lpDNOs this would mean that the body of the 
primary (or, at least where accretion onto it occurs) has acquired a 
rotation period close to that of the belt at its outer edge.

    Perhaps the most direct evidence so far for magnetically 
controlled accretion in a dwarf nova comes from the analysis of 
XMM-Newton observations of OY Car just after an outburst. 
Wheatley \& West (2003) show that the eclipse implies that the 
X-Ray emitting region is noticeably smaller than the surface of the 
primary, and is away from the equator (at latitude $\sim 50^{\circ}$ in their 
diagram). It is the presence of such accretion that may account for 
the rare 48 s QPOs observed in OY Car during quiescence (WWP) 
and the 60 s X-Ray QPOs seen in VW Hyi in quiescence (Pandel, 
Cordova \& Howell 2003). 

    Almost as direct evidence is given by the presence of a variable 
strength inverse P Cygni feature observed in the spectrum of VW 
Hyi during outburst, which has been interpreted as structured 
inflow of gas from the inner edge of a truncated disc (Huang et al.~1996).

\section{THE NATURE OF QPOs}

\begin{figure}
\plotone{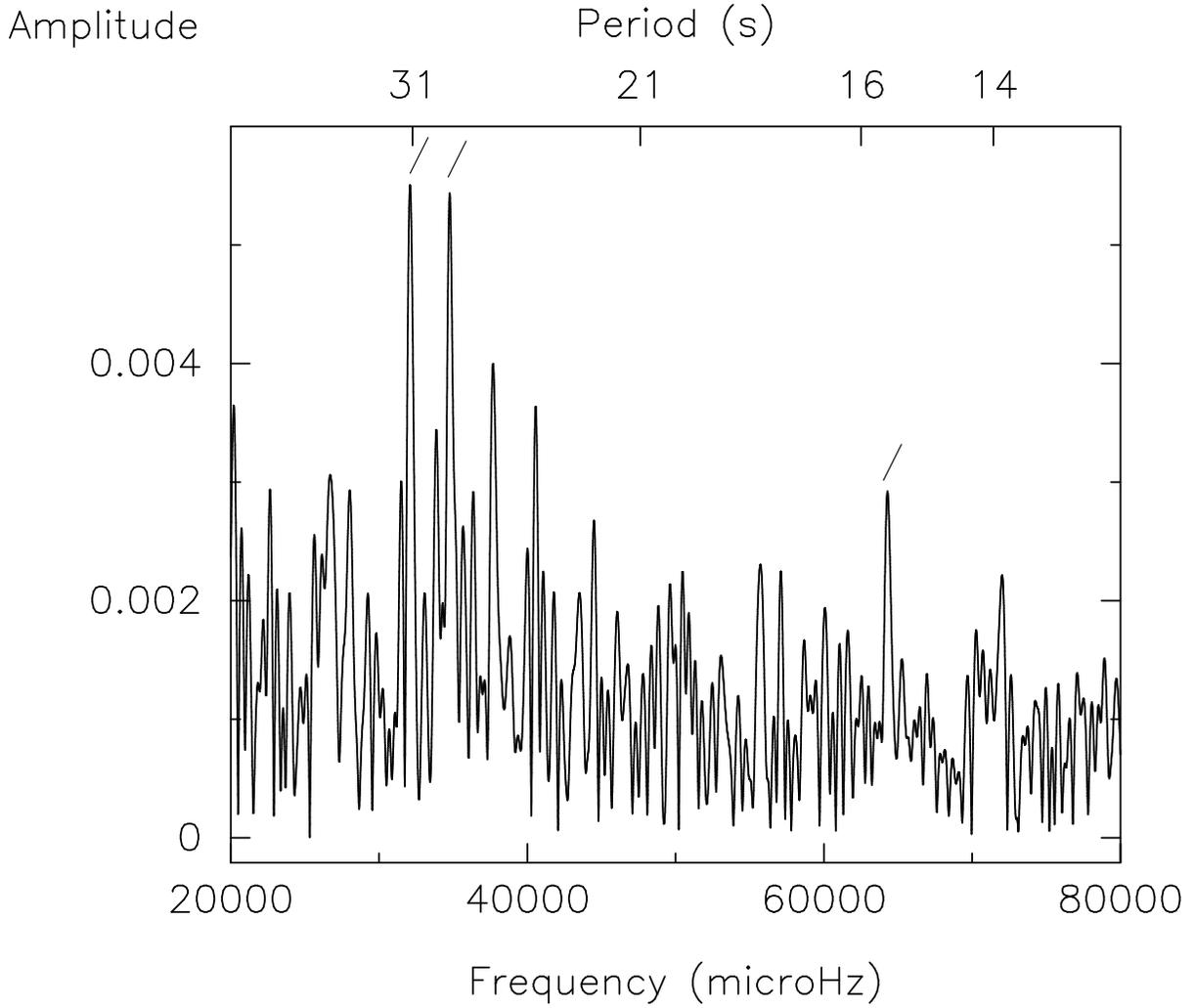}
\caption{Fourier spectrum of a light curve of VW Hyi during outburst showing double DNO and harmonic to the lower frequency
(indicated by bars). From Woudt \& Warner (2002a).}
\label{fig12}
\end{figure}

    The observational evidence available at present shows (WW2b) 
that when a double DNO occurs then it is the shorter of the two 
periods that is at the expected DNO period -- and the profile of that 
signal is sinusoidal. The longer period often has a significant first 
harmonic (e.g., Marsh \& Horne 1998; Fig.~\ref{fig12}). 
The fact that the DNO frequency is accompanied by only one 
sideband shows that this is not a case of modulation of the DNO at 
the QPO frequency; rather it is similar to the dominant sideband 
seen in IPs -- which is caused by `reflection' (more correctly, 
reprocessing) of the rotating beam off the secondary (or other 
structure, such as a thickening of the disc at stream impact, 
revolving at the orbital frequency). By analogy we conclude that 
the longer period DNO component is generated by the rotating 
DNO beam being reprocessed by a structure revolving in a 
prograde direction around the primary at the QPO period. The 
irregular profile of this structure causes the harmonics in the 
reprocessed signal.

    The periods of most QPOs are so short (relative to the orbital
period) that the accretion disc is 
the only possible location for the reprocessing site. If revolving at 
the Keplerian period the site would be in the outer region of the 
disc and would have to have a very large vertical height in order to 
intercept enough of the rotating DNO beam. This is not out of the 
question, but perturbation analyses of accretion discs show that 
there are a number of oscillatory modes available, and that a 
slowly moving prograde travelling wave in the inner disc is the 
most likely to be excited (Lubow \& Pringle 1993). Such a 
travelling wave, excited by a process of magnetic winding and 
reconnection, has been proposed as the structure that generates the 
double DNOs (WW2b). The QPOs themselves are then the result 
of reprocessing and/or obscuration of the radiation from the hot 
inner regions by the travelling wave. 

   The obscuration aspect of QPOs is demonstrated by the 
occasional presence of deep dips at the minima of the QPO 
modulations, which pull the light curve well below its interpolated 
lower envelope (see, e.g., Figure 2 of WW2b). A particularly 
interesting QPO profile is shown by SW UMa, in which the 
minimum of the $\sim$ 370 s modulation is deepened by an apparent 
shallow eclipse of $\sim$ 70 s duration, which is probably a partial 
eclipse of the primary (Fig.~\ref{fig13}).

    This model for optical QPOs -- simple reflection and obscuration 
by an excited mode of the accretion disc -- differs greatly from 
some earlier proposals that also invoke oscillation modes of discs. 
In those (non-magnetic) studies (e.g. Carroll et al.~1985; Collins, 
Helfer \& van Horn 2000) the QPOs are supposed to be oscillations 
of the intrinsic luminosity of the disc itself. This is a viable option 
for non-magnetic primaries, but it remains to be demonstrated 
observationally that such oscillation modes are excited and 
generate detectable QPOs.

    The QPOs observed in the hard X-Ray emission of SS Cyg 
during outburst are not entirely compatible with the above model. 
Although deep dips that appear to be obscurations accompany the 
QPOs (Figure 9 of WMM), the quasi-periodic peaks cannot be due to 
a reflection effect. The overall effect looks more like QPOs of 
$\dot{M}$ at the inner boundary layer. 

\begin{figure}
\plotone{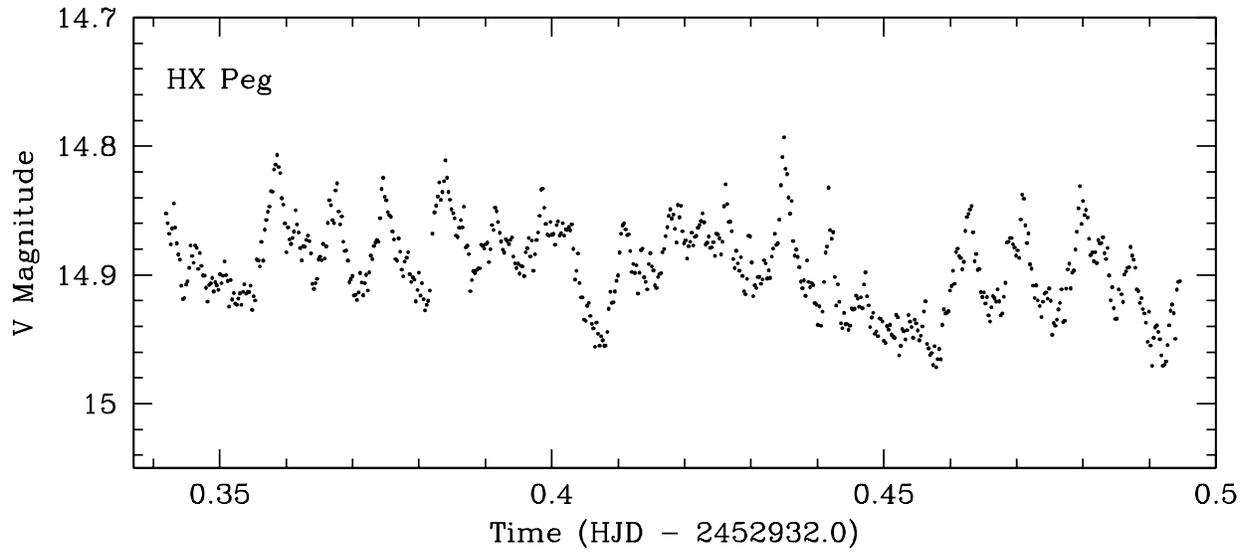}
\caption{Light curve of HX Peg in outburst. The original data have been
binned to 20 s integrations. Courtesy M.L. Pretorius.}
\label{fig12b}
\end{figure}

\begin{figure}
\plotone{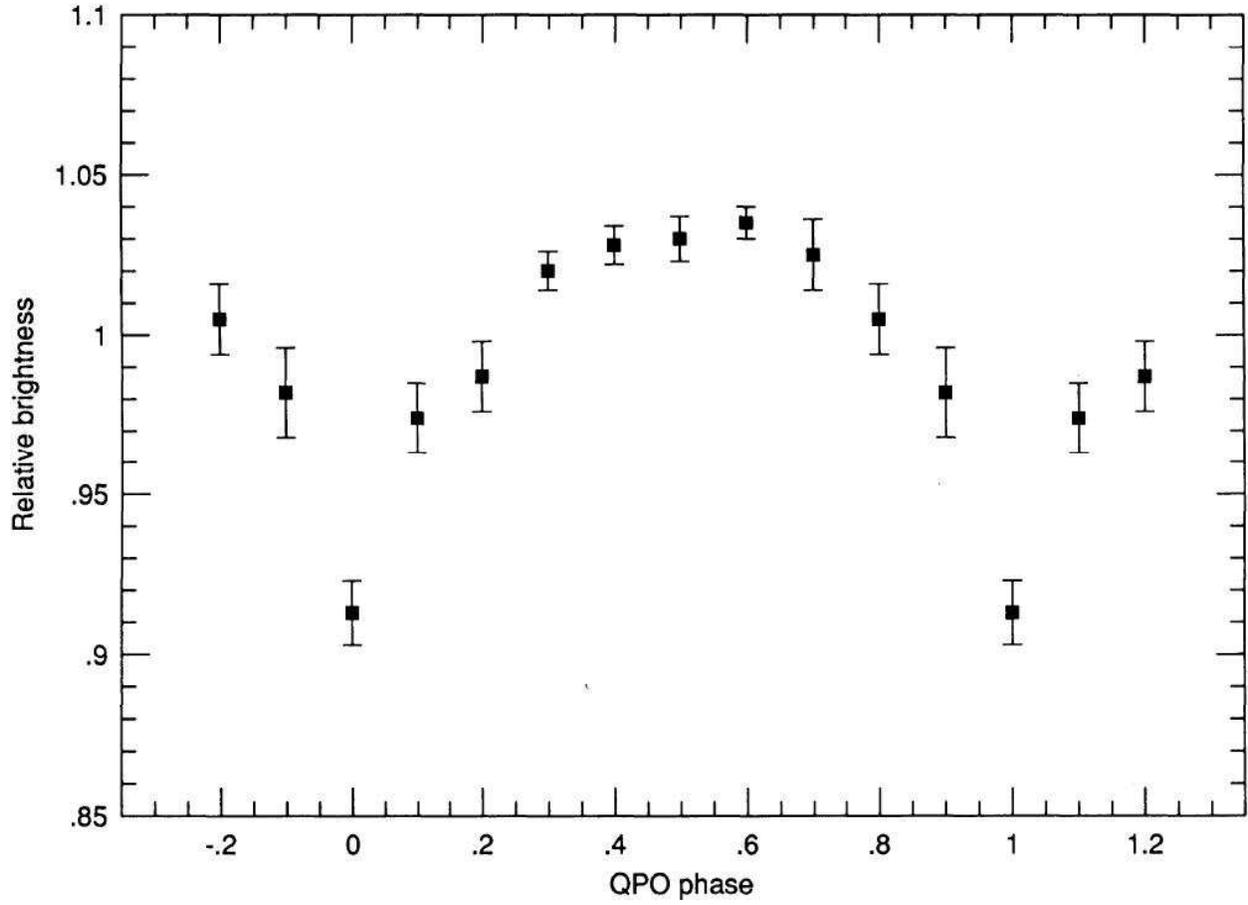}
\caption{The mean light curve of the $\sim$ 370 s QPO in SW UMa (from Kato, Hirata \& Mineshige 1992).
Individual cycles of the QPOs, at higher time resolution, show that the low point is caused by a shallow eclipse.}
\label{fig13}
\end{figure}

    In this regard it is of interest to note parallel studies in other 
fields of astrophysics that have relevance to the QPOs in CVs. Quasi-periodic 
accretion and ejection of dense knots of gas in young 
stellar objects (YSOs) has been modelled as an interaction between 
a magnetic star and its accretion disc. The inner boundary of the 
disc undergoes quasi-periodic radial oscillations, each of which 
strips a ring of gas that falls onto the star (Goodson, B\"ohm \& 
Winglee 1999; Goodson \& Winglee 1999). In the application to 
YSOs, appropriate parameters lead to an estimate of the oscillation 
period as $\sim 100/2{\pi}$ times the Keplerian period at the inner edge of 
the disc. It can be shown that the same mechanism applied to high 
$\dot{M}$ CV discs and $B \sim 10^6$ G produces a similar result, which we 
can interpret as $P_{QPO}/P_{DNO} \sim 100/2{\pi} \sim 16$. Such modulation of 
accretion onto low field CV primaries is therefore another 
mechanism that could produce QPOs at the observed periods. 
These would probably look more like flares than sinusoidal 
modulations. A possible example is shown in Fig.~\ref{fig12b} 
where recurrent flares with a quasi-period of $\sim$ 750 s are seen.
In this light curve, low amplitude ordinary QPOs at 347 s were also
present, demonstrating the permitted co-existence of the two 
types of QPO.

     In similar modelling by Uzdensky (2002) and Uzdensky, 
K\"onigl \& Litwin (2002) of accretion from a disc into the 
magnetosphere of an aligned magnetic rotator, it is found that field 
winding and reconnection can lead to quasi-periodic accretion. The 
first results of 3D modelling of accretion onto an inclined dipole 
rotator have recently appeared in which short time scale QPOs also 
are found (Romanova et al.~2003).

\section{FREQUENCY DOUBLING AND ALIASING}
      
        DNOs in the EUV region of SS Cyg have been observed in 
several outbursts (Mauche 1996b, 1997b, 1998, 2002) and 
constitute the most complete coverage of the evolution of DNOs in 
this wavelength region. The DNOs were found to be mostly 
sinusoidal in profile, except that at the brightest parts of outbursts 
in August 1993 and June 1994 a noticeable first harmonic 
appeared. Then, in the October 1996 outburst, with SS Cyg at its 
brightest, the fundamental disappeared and the signal doubled in 
frequency to become entirely first harmonic, i.e. with a period $\sim$ 3 
s. The next outburst was observed in soft X-Rays and was also 
found to have a 2.8 s DNO (van Teeseling 1997).

   Before discussing possible physical causes of the frequency 
doubling, it is useful to point out that such short oscillation periods 
have implications for optical photometry of DNOs. For example, 
the 9.735 s DNO observed in SS Cyg by Patterson, Robinson and 
Kiplinger (1978) at the unprecedentedly low amplitude of 0.02 
mmag is now seen (as was allowed by the authors at the time) to be 
a beat between a true period of 6.790 s (or its first harmonic) and 
the photometric integration time of 4s. If under-sampled in such a 
way, a sinusoidal signal is reduced in amplitude by a factor of 
${{\sin x}/x}$, where $x = p\,dt/P_{DNO}$ and $dt$ is the integration time. For the 
case under consideration, the true amplitude of the 6.79 s DNO 
would therefore have been 0.04 mmag, or 0.14 mmag if frequency 
doubling had occurred.

   Another interesting example is given by the `type star' for QPOs 
-- namely, RU Peg. The Fourier transform for RU Peg computed by 
Patterson, Robinson \& Nather (1977) shows a QPO centred on 50 s 
and a 0.6 mmag DNO at 11 s; their integration time was 4s. From 
the discussion in Section 4 we would expect DNOs at $\sim$ 50/15 $\sim$ 
3.3.s, not at 11 s. One possibility, therefore, is that the observed 
signal at 11 s was a beat with the integration time, and that the true 
signal was 2.97 s with an amplitude of 2.8 mmag. This opens the 
further possibility that RU Peg had undergone frequency doubling 
at that time. Yet another possibility is that the observed 11 s signal 
was a low amplitude lpDNO and there was no ordinary DNO 
detected at all.

   The lesson to be learnt from this is that it is necessary either to 
carry out photometry of DNOs with integration times of 1 s or less, 
or alternatively to split up the observations using non-
commensurate integration times -- e.g. alternate runs with 4 s and 5 
s integrations.

   The physical cause of frequency doubling is not yet clear -- it is a 
topic to be explored that could lead to greater insight into normal 
DNO behaviour. A transition from single pole to two-pole 
accretion is an obvious possibility. As $\dot{M}$ increases to a 
maximum value the magnetosphere is squashed close to the 
surface of the white dwarf, at which point higher multipole 
components of the magnetic field geometry become more resilient 
than the dipole component (Lamb 1988) and what may have been a 
single visible accretion region could become two or more accreting 
zones (though only one or two are likely to capture most of the 
accretion flow). 

   Another possible model results from the change in viewing 
geometry when the inner radius of the disc approaches the surface 
of the primary. At moderate to high orbital inclinations the primary 
obscures the inner part of the rear of the disc, and the front side of 
the disc, with its accretion curtain, can obscure the lower 
hemisphere and perhaps even part of the equatorial accretion zone. 
Furthermore, accretion columns or curtains are often of lowest 
optical thickness perpendicular to the accretion flow, which 
produces a fan beam. In either case the maximum direct visibility 
of an accretion zone may be near each limb of the rotating primary 
with the result that a single bright region will be seen twice per 
rotation. 

   It would be helpful to have simultaneous EUV/X-Ray and 
optical observations -- the different wavelength regions give 
different points of view, the former showing what happens as seen 
directly from the centre of the system, and the latter showing 
largely what the disc sees. It is possible that frequency doubling 
could occur at the short wavelengths but not at visible 
wavelengths.

    An instructive example of a powerful effect of geometry is given 
by the IP XY Ari, in which the amplitude (in X-Rays) of the 206 s 
periodic pulse is $\sim$ 20\% and is double peaked in quiescence but
$\sim$ 90\% and single peaked in outburst. This is an example of period 
doubling rather than frequency doubling during outburst, the 
interpretation of which is that in quiescence there are two almost 
equal and $180^{\circ}$ out of phase pulses, coming from two-pole 
accretion, but during an outburst the inner edge of the disc moves 
so close to the primary that the lower pole is obscured, removing 
the filling-in effect of the other pole (Hellier, Mukai \& Beardmore 
1997). 

   But there may be other reasons for frequency doubling, 
especially when we note that QPOs can also double (or halve) their 
frequencies. The QPOs in VW Hyi appear to double in frequency 
(WW2b) when the system is approaching quiescence, when the 
inner edge of the disc must be well above the surface of the 
primary and cannot be responsible for any geometric effects. 
Frequency doubling of QPOs in KT Per has been reported 
(Robinson \& Nather 1979). The claimed steady decrease in period 
of QPOs in EF Peg (Kato 2002) is probably more realistically 
interpreted as an approximate frequency doubling near the middle 
of the train of oscillations. These may be understood as switches to 
and from predominantly fundamental and first harmonic 
excitations of the QPO travelling wave, but the cause for this 
remains unknown.

\section{WZ Sge}

     WZ Sge is of particular interest because it seems to be an IP 
while in quiescence and a DNO machine when in outburst -- a 
result of the primary having a magnetic moment almost strong 
enough to anchor the exterior to the interior. I use `almost' because 
in quiescence prior to the 1978 outburst, $P/\dot{P} \sim 1 \times 10^5$ y (see Figure 
7 of Patterson 1980), which is too short a time scale to be a spin-
down of the entire primary. 

\subsubsection{WZ Sge in Quiescence}

        The connection between WZ Sge and DNO/QPO phenomena 
has taken a long time to demonstrate convincingly. Patterson et al.~(1998) 
reviewed the optical observations and added ASCA hard X-Ray 
observations showing modulation at 27.87 s, which are 
characteristic of magnetic channelling onto a white dwarf. HST 
observations also show the dominance of 27.87 s (Skidmore et al.~1999). 
The observed $v \sin i$ (Cheng et al.~1997) for the primary of 
WZ Sge leads to a rotation period 28 $\pm$ 8 s, which is in agreement 
with the magnetic accretor model.

     The presence of an additional persistent 28.952 s modulation, 
which bears no simple relationship to that at 27.87 s, led some 
workers to suggest that it may be a non-radial oscillation of the 
primary (e.g. Robinson, Nather \& Patterson 1978). Patterson 
(1980) originally pointed out that the 28.95 s period could be due 
to reprocessing of a 27.87 s rotating beam from a progradely-moving 
thickening of the disc. Lasota, Kuulkers \& Charles (1999) 
suggested that this reprocessing site could be at the outer rim of the 
accretion disc, moving at the Keplerian period, but WW2b 
suggested, in analogy with their findings in outbursting dwarf 
novae, that the site is near the inner edge of the disc. The existence 
of the hypothesised disc thickening, moving with a revolution 
period of 744 s (the beat period between the two short periods), has 
in fact been directly demonstrated (WW2b). FTs of some of the 
light curves show that there is a 744 s period present (it is probably 
there at all times, but the large variation of amplitude from one 
cycle to the next makes it difficult to detect in the FT) -- in the light 
curve it appears as recurrent dips of variable depth, the most 
prominent of which is around orbital phase 0.25 and has been seen 
in almost all optical curves from the earliest observations 
(Krzeminski \& Kraft 1964). The obscuring source produces the 
deepest dip when it transits across the bright spot as seen from 
behind; it does not transit the bright spot when seen from the front, 
so there is no corresponding deep dip near phase 0.75.

     The infrared (K-band) light curve of WZ Sge is far less variable 
than the light curves at shorter wavelengths and is largely due to 
the modulation caused by viewing the bright spot through an 
optically thin disc (Skidmore et al.~2002). Interestingly, no QPO 
dips are seen in the IR (Ciardi et al.~1998; Skidmore et al.~2002), so 
we deduce that the travelling wave responsible for the QPO dip 
obscures only the hotter central region of the bright spot, and not 
the extended cooler region.

    To be consistent with the interpretation used earlier, we should 
classify the 27.87 s modulation as an lpDNO, related to rotation of 
the main body of the primary, not a DNO. The 27.87/28.95 s pair 
constitutes a double lpDNO. More specifically, WZ Sge is an IP of 
the DQ Her subclass (Warner 1995a), but is of lower Q than 
typical IPs. 

\subsubsection{WZ Sge in Outburst}

    The 2001 superoutburst of WZ Sge provided an opportunity to 
observe how the short period oscillations behave during times of 
greatly increased $\dot{M}$. DNOs were detected only one month after 
the peak of outburst (Knigge et al.~2002; hereafter K2002), and 
later from two to five months after peak brightness (Welsh et al.~2003; 
hereafter W2003). As pointed out by the latter authors, the 
temperature of the primary fell from 29\,000 K to 18\,000 K during 
that time, yet the modulations did not change their character 
appreciably -- this is a characteristic of DNOs rather than white 
dwarf pulsations.

   The earliest observed oscillations (K2002) have periods near 15 s 
and show changes in amplitude and phase that are characteristic of 
DNO behaviour. There were also weak oscillations with periods 
near 6.5 s.  In the later observations (W2003) the 28.96 s 
oscillation was present with some stability, but not the 27.87 s one. 
In addition, there was an 18 s oscillation of lower coherence, 
characteristic of a DNO.

    If we allow that the magnetic field of WZ Sge is barely strong 
enough to couple the white dwarf's exterior to its interior 
(i.e. $B \sim 5 \times 10^5$ G) then we would expect that the very large $\dot{M}$ during 
superoutburst would squash the magnetosphere to the surface of 
the primary. But field enhancement in the equatorial belt could 
allow at least some of the accretion to be magnetically channelled. 
The equatorial belt would start with the spin period of the primary 
itself, i.e. 27.87 s, and would be spun up from there -- perhaps 
producing the 15 s and 18 s DNOs, which appear correlated in the 
usual way with luminosity of the system.  For lpDNOs of 28 s we 
would expect, from the relationships found for other dwarf novae 
(Section 3), DNOs at about one quarter of that period -- i.e. 6 s, 
which is close to the observed 6.5 s DNOs.

   Therefore, phenomenologically, WZ Sge behaves in outburst in 
ways similar to other weakly magnetic dwarf novae. Nevertheless, 
the appearance of the reprocessed signal at 28.96 s so early in the 
outburst decline, which implies the existence of the 744 s 
travelling wave in the model outlines above, requires its excitation 
process and period to be amazingly robust -- and apparently 
independent of the $\dot{M}$ passing through the inner disc.

\section{A CONNECTION WITH X-RAY BINARIES}

   Many of the phenomena exhibited by CVs also appear in X-Ray 
binaries (XRBs), which have neutron stars and black hole (BH) 
candidates accreting from companions. Rapid modulations are 
among the common properties, though in the XRBs the signals 
appear at hard X-Ray energies where the flux is low and until 
recently the oscillations were only detectable in Fourier transforms 
of many thousands of cycles. An overview of QPOs in XRBs is 
given by van der Klis (2002).

\begin{figure}
\plotone{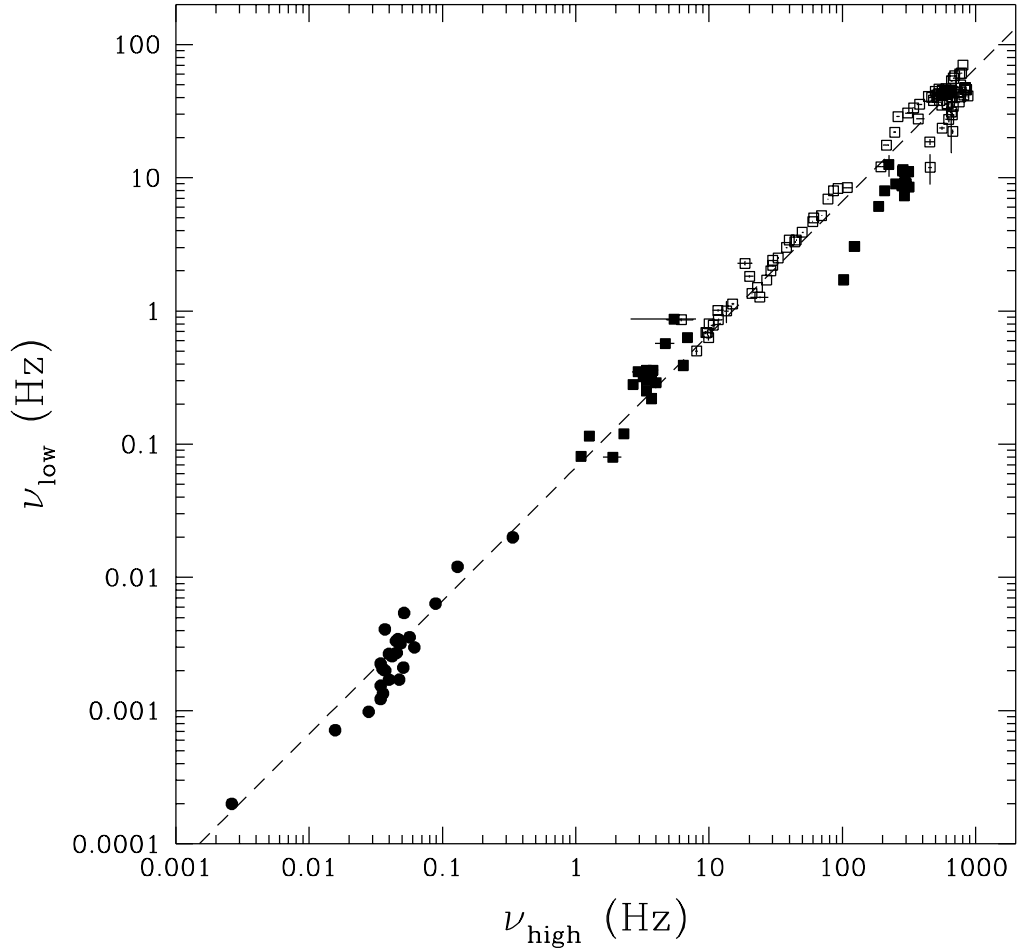}
\caption{The Two-QPO diagram for X-Ray binaries (filled squares: black hole binaries, open squares: neutron star binaries)
and 26 CVs (filled circles). Each CV is only plotted once in this diagram. The X-Ray binary data are from Belloni et al.~(2002) 
and were kindly provided by T.~Belloni. The dashed line marks $P_{QPO}/P_{DNO}$ = 15. From Warner, Woudt \& Pretorius (2003).}
\label{fig14}
\end{figure}

    Some specific similarities to CV behaviour are seen in the phase 
variations in the 5 Hz QPOs in the Rapid Burster (Dotani et al.~1990), 
the double QPOs at kHz frequencies, and the frequency 
dependence on accretion luminosity. But a striking result is that the 
ratio of low to high frequency QPOs that appear in the XRBs is 
close to the value 15 seen in CVs. By considering the CV QPOs 
and DNOs as low and high frequency QPOs we can place them on 
the same diagram as for XRBs -- Fig.~\ref{fig14}. The correlation is seen 
to extend over nearly six orders of magnitude in frequency. This 
does not necessarily mean that exactly the same physics is in 
operation over the entire range, but certainly the same ratio of time 
scales appears everywhere.

    Furthermore, during X-Ray bursts oscillations are often seen 
that lie between the high and low frequency QPOs. They vary in 
frequency and are thought to be caused by hot expanding gas on 
the surface of the neutron star that slips relative to the underlying 
surface (van der Klis 2000). They are closely related to the rotation 
period of the star itself and thus resemble the lpDNOs in CVs 
described in this review.

    A recent important development in the study of X-Ray Binary 
QPOs is an analysis of the signal in the neutron star system 
4U1608-52 (Barret, Olive \& Kluzniak 2003) in which much higher 
time resolution has been achieved. What had been thought to be 
high frequency (800 Hz) QPOs with $Q \sim 10$ have now been shown 
to be short (hundreds of cycles) trains of higher ($Q$ up to 103) 
coherence with a duty cycle $\sim$ 15\% and jumps in frequency of up 
to $\sim$ 0.5\% between the bursts of QPOs. Apart from the implication 
that the QPO amplitudes are typically $\sim$ 8 times what had 
previously been deduced, the high stability for so many cycles 
eliminates many models of X-Ray QPOs (e.g., blobs of accreting 
gas that would be sheared out of existence after only $\sim$ 10 cycles). 
During the phases of high stability the average profile of the QPO 
is closely sinusoidal.

    Consequently the high frequency X-Ray QPOs now look even 
more like CV DNOs -- with period jumps after $\sim$ 100 cycles. 
Among the CVs there are examples of DNOs that are present for 
only part of the time (i.e., a short duty cycle), but also ones (e.g. 
VW Hyi just before quiescence) where high amplitude is 
maintained for at least thousands of cycles.

   The similarities between CV, neutron star and BH rapid 
oscillations may have profound implications. Robertson \& Leiter 
(2003), in gathering evidence that some BHs may have magnetic 
moments, cite these similarities in their compilation. They reason 
that BHs forming from magnetic stars are prevented within a 
Hubble time from reaching their event horizons by the radiation 
pressure that results from pair-creation in the greatly compressed 
and intensified internal magnetic fields. If this is the case, then 
magnetic stars collapsing towards the BH end point behave 
temporarily like over-massive neutron stars, and have more in 
common with magnetic neutron stars and white dwarfs than has 
been previously realised.

\section*{ACKNOWLEDGEMENTS}

    I am greatly indebted to my co-workers in this field -- Dr Patrick 
Woudt, Claire Blackman and Retha Pretorius -- for communicating 
results in advance of publication, and for assistance. Our 
observations have largely been obtained at the Sutherland station 
of the South African Astronomical Observatory. I thank Steve Potter for helpful
comments. My research is 
supported by funds from the University of Cape Town. I 
acknowledge the hospitality of the Nicolaus Copernicus Center in 
Warsaw, where part of this Review was written.

\clearpage

\clearpage

\clearpage

\clearpage

\clearpage

\clearpage

\clearpage

\clearpage

\clearpage

\clearpage

\clearpage

\clearpage

\clearpage

\clearpage

\clearpage

\clearpage

\begin{table}
\begin{center}
\caption{Rapid optical and ultraviolet modulations in cataclysmic variables.}
\begin{tabular}{lllcccl}
\tableline\tableline
Star & Type & $P_{orb}$ & DNO & lpDNO &  QPO & References \\
     &      & (h)       & (s) & (s)   & (s)  &            \\
\tableline
SS Cyg    &  DN  &   6.60  & 6.58$^*$ -- 10.9    & 32 -- 36        &  83-111,730           & 1,2,15,16,53 \\
RU Peg    &  DN  &   8.99  & 11.6 -- 11.8     &$\rightarrow$ ?  &  $\sim$51             &  \\   
VW Hyi    &  NL  &   1.78  & 14.03 -- 40      & $\sim$90        &  400 -- 600           & 3,4,5\\
EM Cyg    &  DN  &   6.98  & 14.6 -- 21.2     &                 &                       &            \\
Z Cam     &  DN  &   6.96  & 16.0 -- 18.8     &                 &                       &        \\
HX Peg    &  DN  &   4.82  & 16.2 -- 16.4     & $\sim$83        &  $\sim$340, 800 -- 1900           & 5,7\\
WX Cet    &  DN  &   1.40  & 17.4             &                 &                       & 61\\
OY Car    &  DN  &   1.50  & 17.6 -- 28.0     & 47.9,116        &  $\sim$320,1500       & 5,6\\
WX Hyi    &  DN  &   1.80  & 19.4             &                 &  $\sim$190,1140,1560  & 5\\
V436 Cen  &  DN  &   1.50  & 19.5 -- 20.1     &                 &   475                 & 4,5\\
HL Aqr    &  NL  &   3.25  & 19.6             &                 &                       &           \\
RR Pic    &   N  &   3.48  & 20 -- 40         &                 &                       &             \\
HT Cas    &  DN  &   1.77  & 20.2 -- 20.4     & $\sim$100       &                       & 1\\
TU Men    &  DN  &   2.82  & 20.6             &                 &   313                 & 7\\
HP Lib    & AMC  &   0.31  &                  &                 &  $\sim$ 290           & 59\\
CR Boo    & AMC  &   0.40  & 21 -- 23         &     62          &  $\sim$300            & 5\\
AQ Eri    &  DN  &   1.46  & 21.0 -- 23.5     & $\sim$90        &  $\sim$280            & 5\\
AH Hya    &  DN  &     -   &    21.55         &                 &                       & 7\\
TY PsA    &  DN  &   2.02  & 21.6, 25.5 -- 30 &  110            &   355                 & 7\\
BR Lup    &  DN  &   1.91  & 21.65            &                 &                       & 7\\
SW UMa    &  DN  &   1.36  & 22.3             &                 &   280 -- 370          & 41\\
KT Per    &  DN  &   3.92  & 22.4 -- 29.3    &$\sim$86,147      &                       &      \\
EC2117    &  NL  &   3.71  & 22.5 -- 25.5     & $\sim$95        &   $\sim$500           & 5\\
WW Cet    &  DN  &   4.22  & 23.1             &    103          &   263                 & 7 \\
SY Cnc    &  DN  &   9.12  & 23.3 -- 33.0     &                 &                       &         \\
VZ Pyx    &  DN  &   1.78  & 23.9             &    112          &   390, $\sim$3000     & 5,29\\
\tableline
\label{tab1}
\end{tabular}
\end{center}
\end{table}

\addtocounter{table}{-1}

\begin{table}
\begin{center}
\caption{Continued: Rapid optical and ultraviolet modulations in cataclysmic variables.}
\begin{tabular}{lllcccl}
\tableline\tableline
Star & Type & $P_{orb}$ & DNO & lpDNO &  QPO & References \\
     &      & (h)       & (s) & (s)   & (s)  &            \\
\tableline
V803 Cen  & AMC  &   0.28  &                  &    176          &                       & 5\\
V1159 Ori &  DN  &   1.50  & 24 -- 34         &    177          &   $\rightarrow$ ?     & 7,42\\
AH Her    &  DN  &   5.93  & 24.0 -- 38.8     & $\sim$100       &                       & 1\\
CN Ori    &  DN  &   3.91  & 24.3 -- 32.6     &                 &                       & 5\\
IX Vel    &   N  &   4.65  & 24.6 -- 29.1     &                 &   $\sim$500           & 25\\
U Gem     &  DN  &   4.25  & $\sim$25         & $\sim$146       &                       & 26\\
Z Cha     &  DN  &   1.79  & 25.1 -- 27.7     &                 &   585                 & 5\\
V893 Sco  &  DN  &   1.82  & 25.2             &                 &   $\sim$350           & 5,30\\
BP Lyn    &  NL  &   3.67  & 25.5             &                 &                       &         \\
AM CVn    & AMC  &   0.28  & 26.3             &                 &   290,820             & 55,56,57,58\\
WZ Sge    &  DN  &   1.36  & 27.87, 28.95     &                 &   742                 & 4,8,9,10,11,12,13,\\
          &      &         & 14.48 + others   &                 &                       & 14,31,45,46,47\\
V2051 Oph &  DN  &   1.50  & 28.06, 29.77     &                 &   486,1800            & 4,18   \\
          &      &         &  42.2            &                 &                       &    \\
UX UMa    &  NL  &   4.72  & 28.5 -- 30.0     &                 &   $\sim$650           & 3,17\\
V3885 Sgr &  NL  &   4.94  & 29 -- 32         &                 &                       &  \\
HP Nor    &  DN  &    -    & 35.2             &                 &                       & 7\\
RX And    &  DN  &   5.08  & 36               &                 &   $\sim$1000          & 5\\
BP Cra    &  DN  &    -    & 38.6             &                 &                       & 7\\
V436 Car  &  NL  &   4.21  & $\sim$40         &      123        &                       & 38\\
V533 Her  &   N  &   3.52  & 63.63            &                 &     1400              & 5,28            \\
YZ Cnc    &  DN  &   2.08  &                  & $\sim$90        &                       &         \\
LX Ser    &  NL  &   3.80  & $\sim$140        & $\rightarrow$ ? &                       &       \\
X Leo     &  DN  &   3.95  &  89 -- 160       &                 &                       & 7 \\
TW Vir    &  DN  &   4.38  & 112 -- 121       &                 &    1000               & 7                     \\
V373 Sct  &   N  &    -    & 258.3            & $\rightarrow$ ? &                       & 37\\
\tableline
\label{tab1b}
\end{tabular}
\end{center}
\end{table}

\addtocounter{table}{-1}
\begin{table}
\begin{center}
\caption{Continued: Rapid optical and ultraviolet modulations in cataclysmic variables.}
\begin{tabular}{lllcccl}
\tableline\tableline
Star & Type & $P_{orb}$ & DNO & lpDNO &  QPO & References \\
     &      & (h)       & (s) & (s)   & (s)  &            \\
\tableline
BT Mon    &   N  &   8.01  &                  &                 &   $\sim$1800          & 5,27\\
GK Per    &   N  &   1.99d & 360 -- 380       &                 &     5000              & 19,20,21                    \\
KR Aur    &  DN  &   3.91  &                  &                 &    400 -- 900         & 22\\
EF Peg    &  NL  &   1.92  &                  &                 &    400, 1080          & 39\\
TV Crv    &  DN  &   1.50  &                  &                 &    600                & 49\\
RW Sex    &  NL  &   5.93  &                  &                 &    620, 1280          &    \\
V842 Cen &    N  &     -   &                  &                 &    750 -- 1300        &   37\\
EC0528-58 &  NL  &     -   &                  &                 &    900 -- 1560        &   48\\
TT Ari    &  NL  &   3.30  &                  &                 &    900 -- 1600        &  33,50,51\\
V442 Cen  &  DN  &    -    &                  &                 &    925                &  \\
V442 Oph  &   NL &    2.98 &                  &                 &     1000              &     50\\
RXJ 1643  &  NL  &   2.89  &                  &                 &    1000               &    50\\
BH Lyn    &  NL  &   3.74  &                  &                 &    1030               &    50\\
AH Men    &  NL  &   3.01  &                  &                 &    1100               &    50\\
V795 Her  &  NL  &   2.60  &                  &                 &    1150               & 32,50\\
WX Ari    &  NL  &   3.34  &                  &                 &    1180               &   50\\
SS Aur    &  DN  &   4.39  &                  &                 &    1200 -- 1800       &     52\\
V751 Cyg  &  NL  &   3.47  &                  &                 &    1230               &   34 \\
LS Peg    &  NL  &   4.20  &                  &                 &    1240               &  35,36,40\\
NSV 10934 &  DN  &   1.7   &                  &                 &    1300               & 60\\
V426 Oph  &  DN  &   6.85  &                  &                 &    1680               &       \\
GO Com    &  DN  &   1.58  &                  &                 &    1980               &   \\
V592 Cas  &  DN  &   2.76  &                  &                 &    2160               &44 \\
CW Mon    &  DN  &   4.23  &                  &                 &    2200               &54 \\
SU UMa    &  DN  &   1.83  &                  &                 &    2280               &      \\
MV Lyr    &  NL  &   3.20  &                  &                 &    $\sim$3000         &    23,24\\
\tableline
\label{tab1c}
\end{tabular}
\end{center}
\end{table}

\clearpage
{\footnotesize 
Notes Table 1: DN = dwarf nova, N = nova, NL = nova-like, AMC = AM CVn star, $^*$ A 
period approximately half this was also observed.\newline
The References are additional to those given in Table 8.2 of Warner (1995).\newline
1 Patterson 1981; 2 Mauche \& Robinson 2001; 3 Woudt \& Warner 2002a; 4 
Warner \& Woudt 2002; 5 Warner, Woudt \& Pretorius 2003; 6 Marsh \& 
Horne 1998; 7 M. L. Pretorius (unpublished); 8 Knigge et al.~2002; 9 
Skidmore et al.~2002; 10 Skidmore et al.~1999; 11 Patterson et al.~1998; 12 
Welsh et al.~1997; 13 Skidmore et al.~1997; 14 Provencal \& Nather 1997; 15 
Mauche 1996a; 16 Mauche 1997a; 17 Knigge et al.~1998; 18 Steeghs et al.~2001; 
19 Nogami, Kato \& Baba 2002; 20 Morales-Rueda, Still \& Roche 
1996; 21 Morales-Rueda, Still \& Roche 1999; 22 Kato et al.~2002; 23 
Pavlenko \& Shugarov 1999; 24 Kraicheva et al.~1999a; 25 Williams \& 
Hiltner 1984; 26 Long et al.~1996; 27 Smith, Dhillon \& Marsh 1998; 28 
Rodriguez-Gil \& Martinez-Pais 2002; 29 Remillard et al.~1994; 30 Bruch, 
Steiner \& Gneiding 2000; 31 Patterson et al.~2002a; 32 Rodriguez-Gil et al.~2002; 
33 Kraicheva et al.~1999b; 34 Patterson et al.~2001; 35 Taylor et al.~1999; 
36 Rodriguez-Gil et al.~2001; 37 Woudt \& Warner 2003; 38 Woudt \& 
Warner 2002b; 39 Kato 2002; 40 Szkody et al.~2001; 41 Nogami et al.~1998; 
42 Patterson et al.~1995. 43 Skillman, Patterson \& Thorstensen 1995; 44 
Kato \& Starkey 2002; 45 Provencal \& Nather 1997; 46 Araujo-Betancor 
2003; 47 Welsh et al.~2003; 48 Chen et al.~2001; 49 Uemura et al.~2001; 50 
Patterson et al.~2002b; 51 Tremko et al.~1996; 52 Tovmassian 1988; 53 
Mauche 2002; 54 Kato et al.~2003; 55 Patterson et al.~1979; 56 Skillman et al.~1999; 
57 Patterson et al.~1992; 58 Provencal et al.~1995; 59 Patterson et al.~2002c; 
60 Kato et al.~2004; 61 J. Patterson (private communication).  
}

\clearpage

\begin{table}
\begin{center}
\caption{Rapid oscillations in X-Rays.}
\begin{tabular}{lllcccl}
\tableline\tableline
Star & Type & $P_{orb}$ & Periods & Energy & State & References \\
     &      & (h)       & (s) &   &   &            \\
\tableline
SS Cyg  &    DN    &  6.60   &    7.4 -- 10.7        &    Soft &     O  &  1-3,9\\
        &          &         &       2.8$^*$         &    Soft &     O  &     10\\
        &          &         &    155 -- 245         &    Hard &     O  &     17\\
VW Hyi  &    DN    &  1.78   &   14.06, 14.2-14.4    &    Soft &     O  &     6   \\
        &          &         &    63 -- 68           &    Soft &     O  &     6   \\
        &          &         &    $\sim$60           &    Hard &     Q  &    14\\
        &          &         &    $\sim$500          &    Hard &     O  &    12 \\
HT Cas  &    DN    &  1.77   &      21.85:           &    Hard &     Q  &    4,7\\
U Gem   &    DN    &  4.25   &      25-29            &    Soft &     O  &    2,8\\
        &          &         &     121,135           &    Hard &     Q  &     7\\
        &          &         &       585             &    Hard &     O  &     7\\
WZ Sge  &    DN    &  1.36   &       27.8            &    Hard &    Q   &    13\\
SU UMa  &    DN    &  1.83   &       33.93:          &    Hard &     Q  &     4\\
YZ Cnc  &    DN    &  2.21   &        222            &    Hard &     Q  &     7\\
RW Sex  &    NL    &  5.93   &        254            &    Hard &        &       7\\
AB Dra  &    DN    &  3.65   &       290             &    Hard &     O  &     7\\
OY Car  &    DN    &  1.50   &      2240             &    Soft &      Q &     11\\
GK Per  &   N, DN  &  1.99d  &   3000-5000           &    Hard &     O  &  15,16\\
\tableline
\label{tab2}
\end{tabular}
\end{center}
{\footnotesize 
Notes Table 2: O = outburst, Q = quiescence. A colon (:) denotes a less certain 
observation. $^*$ Frequency doubling had occurred.\newline
1 Cordova et al.~1980; 2 Cordova et al.~1984; 3 Watson, King \& 
Heise 1985; 4 Eracleous, Patterson \& Halpern 1991; 5 Jensen et al.~1983; 
6 van der Woerd, Heise \& Bateson 1986; 7 Cordova \& 
Mason 1984; 8 Mason et al.~1988; 9 Jones \& Watson 1992; 10 van 
Teeseling 1997; 11 Ramsay et al.~2001; 12 Wheatley et al.~1996; 12 
Wheatley et al.~1996; 13 Patterson et al.~1998; 14 Pandel, Cordova 
\& Howell 2003; 15 Watson, King \& Osborne 1985; 16 Ishida et al.~1996; 
17 Wheatley, Mauche \& Mattei 2003.
}

\end{table}

\end{document}